\RequirePackage{fix-cm}
\documentclass[smallextended]{svjour3}       % onecolumn (second format)
\smartqed  % flush rntrinsight qed marks, e.g. at end of proof
\usepackage{cite}
\usepackage{appendix}
\usepackage{amsmath}
\usepackage{lineno}

\newcommand{\bi}{\begin{itemize}[leftmargin=0.4cm]}
	\newcommand{\ei}{\end{itemize}}
\newcommand{\be}{\begin{enumerate}[leftmargin=0.4cm]}
	\newcommand{\ee}{\end{enumerate}}
\usepackage{hyperref}
\usepackage{booktabs}
\usepackage{multirow}
\usepackage[table,xcdraw]{xcolor}
\usepackage{dblfloatfix}

\usepackage[ruled,vlined]{algorithm2e}
\usepackage{algorithmic}

\usepackage{xfakebold}

\newcommand{\fbseries}{\unskip\setBold\aftergroup\unsetBold\aftergroup\ignorespaces}
\makeatletter
\newcommand{\setBoldness}[1]{\def\fake@bold{#1}}
\makeatother
\usepackage{adjustbox}
\usepackage{bigstrut}
\usepackage{wrapfig}
\usepackage{adjustbox}
\usepackage{enumitem}
\usepackage[skins]{tcolorbox}
\usepackage{subfigure}

\usepackage{color}
\usepackage{balance}
\usepackage[usestackEOL]{stackengine}
\usepackage{xcolor}
\usepackage{graphicx}

\newcommand{\tion}[1]{\S\ref{tion:#1}}

\newcommand{\tbl}[1]{Table~\ref{tbl:#1}}

\begin{document}

\makeatletter
\makeatother
\newenvironment{RQ}{\vspace{2mm}\begin{tcolorbox}[enhanced,width=\linewidth,size=fbox,fontupper=\normalsize,colback=red!5,drop shadow southeast,sharp corners]}{\end{tcolorbox}}

\title{Learning to Recognize Actionable \\Static Code Warnings (is Intrinsically Easy)}
% \title{Are Deep Neural Networks the Best Choice\\ for Reasoning about SE?}

\author{Xueqi Yang \and Jianfeng Chen \and Rahul Yedida \and Zhe Yu \and Tim Menzies}
%\authorrunning{Short form of author list} % if too long for running head

\institute{Department of Computer Science, North Carolina State University, Raleigh, NC, USA \\
\email{xyang37@ncsu.edu, jchen37@ncsu.edu, ryedida@ncsu.edu, zyu9@ncsu.edu,\newline
$\spadesuit$ Corresponding author:
timm@ieee.org}             \\
        %     \emph{Present address:} of F. Author  %  if needed
        %   \and
        %   S. Author \at
        %       second address
}

\date{Received: date / Accepted: date}

\maketitle

\begin{abstract}

Static code warning tools often generate warnings that programmers ignore. Such tools can be made more useful via data mining algorithms that select the ``actionable'' warnings; i.e. the warnings that are usually not ignored.~\\
In this paper, we look for actionable warnings within a sample of  5,675 actionable warnings seen in 31,058 static code warnings from FindBugs. We find that data mining algorithms can find actionable
warnings with remarkable ease. Specifically, a range of data mining methods (deep learners, random forests, decision tree learners, and support vector machines) all 
achieved very good results
(recalls and AUC(TRN, TPR) measures usually over 95\%  and  false alarms usually under 5\%).~\\
Given that all these learners succeeded so easily, it is appropriate  to ask if there
is something about this task that is inherently easy. We report that while our data sets have up to 58 raw features, those features can be approximated by less than two underlying dimensions. For such intrinsically simple data, many different kinds of learners can generate useful models with similar performance.~\\
Based on the above, we conclude that
learning to recognize 
actionable
static code warnings is easy, using a
wide range of learning algorithms, since the underlying data is intrinsically simple.
If we had to pick one particular learner for this task, we would suggest linear SVMs (since, at least in our sample, that learner ran relatively quickly and achieved the best median  performance) and we would not  recommend deep learning  (since this data is intrinsically very simple).

% Based on these results, we have three recommendations. Firstly, it is possible and effective to augment static code warning tools with a post-processor that prune away the  warnings that programmers will ignore. Secondly, always assess data mining performance by applying some variations to the test data.  Thirdly, before selecting a data mining algorithm, always check the intrinsic dimensionality of the data.

% \PACS{PACS code1 \and PACS code2 \and more}
% \subclass{MSC code1 \and MSC code2 \and more}
\end{abstract}

\section{Introduction}
\label{sec:introduction}

\begin{figure}[!b]
 \begin{center}
\includegraphics[width=4in]{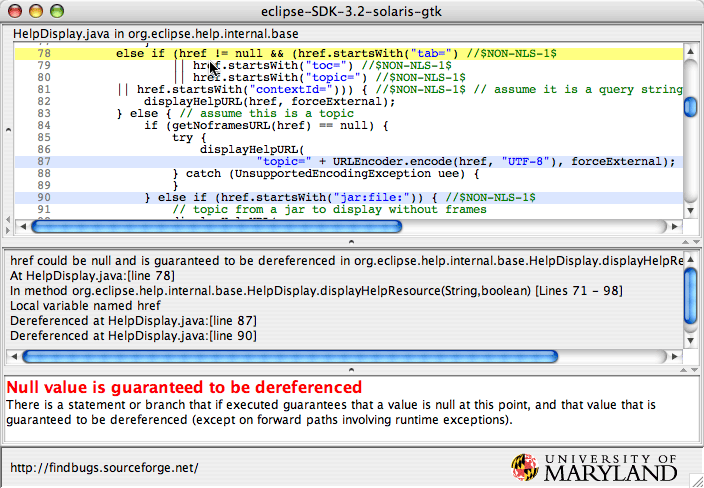} 
\end{center}
\caption{Static code analysis   and  FindBugs. From 
\url{http://findbugs.sourceforge.net/}.}
\label{fig:fb}
\end{figure}
% This paper is about
% (a)~recognizing which static code warnings can be safely ignore; and (b)~what kinds of algorithms are best/worse for that task.
% We will find that:
% \bi
% \item
% We can automatically learn, with a high degree of recall and AUC with very low false alarm, which warnings can be ignore;
% \item
% A certain class of very popular algorithms (collectively called ``deep learning'')
% are very bad for this task-- for reasons that are fundamental to the nature of static warning data.
% \ei
Static code warnings comment on a range of potential defects such as common programming errors, code styling, in-line comments common programming anti-patterns, style violations, and questionable coding decisions~\cite{ayewah2008using}.
Static code warning tools are quite popular. For example the FindBugs static code analysis tool (shown in Figure~\ref{fig:fb})
has been downloaded over a million times.

One issue with static code warnings is that they generate a large number of false positives.
Many programmers routinely ignore most of the static code warnings, finding them irrelevant or spurious~\cite{wang2018there}. 
Such warnings are considered as ``unactionable'' since programmers never take action on them.
Between 35\% and 91\% of the warnings generated from static analysis tools are known to be unactionable~\cite{heckman2011systematic}.
Hence it is prudent to learn to recognize what kinds of warnings programmers usually act upon. With such a classifier, static code warning tools can be made more useful by first pruning away the unactionable warnings.

As shown in this paper, data mining methods can be used to generate very accurate models for this task.
This paper searches for 5,675 actionable warnings within a sample of 31,058 static code warnings generated by FindBugs on nine open-source Java projects~\cite{wang2018there}.
After the experiment (where we trained on release $i$ then tested on release $i+1$), we built models (using linear SVM) that predicted for actionable warnings with recalls over
87\%; false alarms under 7\%; and AUC over 97\%. These results are a new high watermark in this area of research since they outperform a prior state-of-the-art result (the so-called ``golden set'' approach reported at ESEM'18 by Wang et al.~\cite{wang2018there}).

%https://docs.google.com/spreadsheets/d/1BQHsctwv_00eaUJ2kZU2BheJJpAwA8htUd_bKYCoQyI/edit#gid=0

Given that all these learners succeeded so easily, it is appropriate to ask if there
is something about this task that is inherently easy. We report that while our data sets have up to 58 raw features, those features can be approximated by less than two underlying dimensions. For such intrinsically simple data, many different kinds of learners can generate useful models with similar performance.
Hence, at least for this task, complex methods like deep learners ran far slower
and performed no better than much simpler methods. 
This was somewhat surprising since, to say the least,
there are many advocates of deep learning for
software analytics (e.g.~\cite{lin2018sentiment,guo2017semantically,chen2019mining,gu2016deep,nguyen2017exploring,choetkiertikul2018deep,zhao2018deepsim,white2016deep}).

The rest of this paper is structured as follows. The background to this work is introduced in Section~\ref{sec:background}. 
Our methodology is described in Section~\ref{sec:method}. In Section~\ref{sec:experiment} and Section~\ref{sec:intrinsic}, we analyse   experiment results.
Threats to validity and future work are discussed in Section~\ref{sec:discussion}. Our conclusions, drawn in Section~\ref{sec:conclusion}, will be three-fold:
\be
\item
It is possible and effective to augment static code warning tools with a post-processor that prunes away the warnings that programmers will ignore. 
 \item
Before selecting a data mining algorithm, always check the intrinsic dimensionality of the data.
\item After checking the intrinsic dimensionality, match the complexity of the learner to the complexity of the problem.
\ee
To facilitate other researchers in this area, all our scripts and data are freely available on-line\footnote{ \url{https://github.com/XueqiYang/intrinsic_dimension}.}.

\section{Background}
\label{sec:background}

\subsection{Studying Static Code Warnings}

Static code warning tools detect potential static code defects in source code or executable files at the stage of software product development. The distinguishing feature of these tools is that they make their comments without reference to a particular input. Nor do they use feedback from any execution of the code being studied.
Examples of these tools include PMD\footnote{\url{https://pmd.github.io/latest/index.html}}, Checkstyle\footnote{\url{https://checkstyle.sourceforge.io/}} and the  FindBugs\footnote{\url{http://findbugs.sourceforge.net}} tool featured in Figure~\ref{fig:fb}.

As mentioned in the introduction, previous research work shows that 35\% to 91\% warnings reported as bugs by static warning analysis tools can be ignored by programmers~\cite{heckman2011systematic}.
% Such SA tools such as FindBugs utilizes static analysis (SA) techniques to inspect source code or executable files for the occurrence of bug patterns (i.e., the code idiom that is often an error). These bugs detected by FindBugs are grouped into a pattern list, (i.e, performance, style, correctness and so forth) and each bug is reported by FindBugs with priority from 1 to 20 to measure the severity, which is finally grouped into four scales either scariest, scary, troubling, and of concern.
% However, false positives caused by excessive warning generation and high portion of unactionable alerts (which developers cannot act on) reported by static analysis tools are making developers reluctant to use these tools. Previous research observes that 35\% to 91\% of warnings generated by static warning analysis tools are actually unactionable~\cite{heckman2011systematic}. 
% % Moreover, 
This high false alarm rate is one of the most significant barriers for developers to use these tools~\cite{thung2015extent,avgustinov2015tracking,johnson2013don}.  
%Heckman et al.~\cite{heckman2011systematic,heckman2013comparative} performed a systematic study in the related literature that focus on classifying or prioritizing alters generated by static analysis tools. 
Various approaches have been tried to reduce these false alarms including graph theory~\cite{boogerd2008assessing,bhattacharya2012graph}, statistical models~\cite{chen2005novel},
% Wang et al. \cite{wang2018there} aims at exploring whether there is a golden feature set for actionable warning identification by conducting a systematic experimental evaluation of all the public available features. The major goal is to kill the high false positive rate of the reported warnings using reduced feature dimensions.  Yuksel et al.~\cite{yuksel2013automated} evaluated 34 machine learning algorithms using 10 software characteristics to classify static warnings. Another work from Ruthruff et al.~\cite{ruthruff2008predicting} applied logistic regression models to predict actionable warnings.
% Avgustinov et al. \cite{avgustinov2015tracking} presented
% an approach to track static analysis warnings over the history of
% a project, and further use the information to capture developer’s
% characteristics. Smith et al. \cite{smith2015questions} investigated questions developers asked while diagnosing potential security vulnerabilities with static analysis. Thung et al. \cite{thung2015extent} studied to what extent could field defects be detected by the state-of-the-art static analysis tools.
and ranking schemes~\cite{kremenek2004correlation}. For example, Allier et al.~\cite{allier2012framework} proposed a framework to compare 6 warning ranking algorithms and identified the best algorithms to rank warnings. Similarly, Shen et al.~\cite{shen2011efindbugs} employed a ranking technique to sort true error reports before anything else.  Some other works also prioritize warnings by dividing the results into different categories of impact factors~\cite{liang2010automatic} or by analyzing software history~\cite{kim2007prioritizing}.
\begin{table}[]
\footnotesize
% \textwidth
% \linewidth
\caption{Categories of Wang et al.~\cite{wang2018there}'s selected features.~(8 categories are shown in the left column, and 95 features explored in Wang et al. are shown in the right column with 23 golden features in bold.)}
\tabcolsep=0.11cm
% \begin{adjustbox}{max width=0.48\textwidth}
\begin{center}
\begin{tabular}{ll}
% \begin{tabular}{l|l}
\hline
\textbf{Category}  &   \textbf{Features} \\ \hline
Warning combination & \begin{tabular}[c]{@{}l@{}}size content for warning type;\\ size context in method, file, package;\\ \textbf{warning context in method, file,} package;\\ \textbf{warning context for warning type};\\ fix, non-fix change removal rate;\\ \textbf{defect likelihood for warning pattern};\\ variance of likelihood;\\ defect likelihood for warning type;\\ \textbf{discretization of defect likelihood}; \\ \textbf{average lifetime for warning type};\end{tabular} \\ \hline
Code characteristics & \begin{tabular}[c]{@{}l@{}}method, file, package size;\\ comment length;\\ \textbf{comment-code ratio};\\ \textbf{method, file depth};\\ method callers, callees;\\ \textbf{methods in file}, package;\\ classes in file, \textbf{package};\\ indentation;\\ complexity;\end{tabular} \\ \hline
Warning characteristics & \begin{tabular}[c]{@{}l@{}}\textbf{warning pattern, type, priority,} rank;\\ warnings in method, file, \textbf{package};\end{tabular} \\ \hline
File history & \begin{tabular}[c]{@{}l@{}}latest file, package modification;\\ file, package staleness;\\ \textbf{file age}; \textbf{file creation};\\ deletion revision; \textbf{developers};\end{tabular} \\ \hline
Code analysis & \begin{tabular}[c]{@{}l@{}}call name, class, \textbf{parameter signature},\\ return type; \\ new type, new concrete type;\\operator;\\ field access class, field; \\catch;\\ field name, type, visibility, is static/final;\\ \textbf{method visibility}, return type,\\ is static/ final/ abstract/ protected;\\ class visibility, \\ is abstract / interfact / array class;\end{tabular} \\ \hline
Code history & \begin{tabular}[c]{@{}l@{}}added, changed, deleted, growth, total, percentage \\ of LOC in file in the past 3 months;\\ \textbf{added}, changed, deleted, growth, total, percentage \\ of LOC in file in the last 25 revisions;\\ \textbf{added}, changed, deleted, growth, total, percentage \\ of LOC in package in the past 3 months;\\ added, changed, deleted, growth, total, percentage \\ of LOC in package in the last 25 revisions;\end{tabular} \\ \hline
Warning history & \begin{tabular}[c]{@{}l@{}}warning modifications;\\ warning open revision;\\ \textbf{warning lifetime by revision}, by time;\end{tabular} \\ \hline
File characteristics & \begin{tabular}[c]{@{}l@{}}file type;\\ file name; \\package name;\end{tabular} \\ \hline
\end{tabular}
\end{center}
\label{table:variables}
% \end{adjustbox}
\end{table}

Another approach, and the one taken by this paper, utilizes machine learning algorithms to recognizing which  static code warnings that programmers will act upon~\cite{wang2016automatically,shivaji2009reducing,hanam2014finding}.  For example, when Heckaman et al. applied 15   learning algorithms to 51 features derived from static analysis tool,
they achieved  recalls of  83-99 \% (average across 15 data sets)~\cite{heckman2009model}.

% Several researchers propose to combine machine learning techniques to help developers identify actionable warnings, e.g., finding alerts with similar code patterns and building prediction models to classify new alerts~\cite{hanam2014finding}.

% Heckman et al. identified set of alert characteristics and built models to classify actionable and unactionable alerts~~\cite{heckman2009model}. 

% \textcolor{red}{A conclusion drawn that XXX is the pior state of the art.}

\subsection{Wang et al.'s ``Golden Set''}
\label{tion:dataset}
The data for this paper comes from a recent study by Wang et al.~\cite{wang2018there}.
They conducted a systematic literature review to collect all public available static code features generated by widely-used static code warning tools (116 in total):
\bi
\item
All the values of these collected features were extracted from warning reports generated by FindBugs based on 60 revisions of 12 projects. These projects are selected due to their source code
histories spanning for multiple years, sufficient project size and version control system to extract features from~\cite{wang2018there}.

\item To ensure that the difference between prior and later revision intervals of a project are adequate to draw solid conclusions, Wang et al.~\cite{wang2018there} set revision intervals for different projects, e.g., 3 months for Lucene and 6 months for Mvn. Each project in this study has at least two-years commit history.
\item
Six machine learning classifiers were then employed to automatically identify actionable static warning
(random forests, decision trees, a boosting algorithm, naive bayes, linear regression, and support vector machines).
\item
After applying a greedy backward selection algorithm to eliminate noneffective features to the results of those learners, they isolated 23 features as the most useful ones for identifying actionable warnings. 
\item
They called these features the ``golden set''; i.e. the features most important for recognizing actionable static code warnings.
\ei
To the best of our knowledge, this is the most exhaustive research about static warning characteristics yet published. Therefore, we encourage other researchers in the SE community to continue exploring this data as a standard basis due to the exhaustive feature construction process and the long-spinning code history.

% Please add the following required packages to your document preamble:
% \usepackage{booktabs}
% \usepackage{multirow}
\begin{table}[!t]
\caption{Summary of data distribution.}
% \small
\footnotesize
\begin{center}
\begin{tabular}{@{}rrrrrrr@{}}
\midrule
 &&  & \multicolumn{2}{c}{training set} & \multicolumn{2}{c}{test set} \\ \midrule
&Dataset &  Features& 
 \begin{tabular}[c]{@{}c@{}}instance \\ counts\end{tabular} & \begin{tabular}[c]{@{}c@{}}actionable \\ ratio(\%)\end{tabular} & \begin{tabular}[c]{@{}c@{}}instance \\ counts\end{tabular} & \begin{tabular}[c]{@{}c@{}}actionable \\ ratio(\%)\end{tabular} \\ \midrule
 
 & commons &39& 725 & 7 & 786 & 5 \\ 
 & phoenix &44&  2235 & 18 & 2389 & 14 \\
 & mvn (maven) &47& 813 & 8 & 818 & 3 \\
 & jmeter &49& 604 & 25 & 613 & 24 \\
 & cass (cassandra) &55& 2584 & 15 & 2601 & 14 \\
 & ant &56& 1229 & 19 & 1115 & 5 \\
 & lucence &57& 3259 & 37 & 3425 & 34 \\
 
 & derby &58& 2479 & 9 & 2507 & 5 \\
 
 & tomcat &60& 1435 & 28 & 1441 & 23 \\
 \bottomrule
\end{tabular}
\end{center}
\label{table:distribution}
\end{table}

As shown in Table~\ref{table:variables},
the  ``golden set'' features   fall into eight categories.
These features are the independent variables used in this study.

To assign dependent labels, we applied the methods of
Liang et al.~\cite{liang2010automatic}.
They defined a specific warning as actionable if it is closed after the later revision interval.
Otherwise, it is labeled as unactionable.
Also, after Liang et al., anything labeled a ``minor alert'' is deleted and ignored.

By analyzing FindBugs output from two consecutive releases of nine software projects, collecting the features of Table~\ref{table:variables}, and then applying the Liang et al. definitions, we created the data of
Table \ref{table:distribution}. 
In this table, the ``training set'' refers to release  $i-1$ and the ``test set'' refers to release $i$. In this study, we only employ two latest releases.

Note that, for any particular data set,
the 23 categories of  Table~\ref{table:variables},
can grow to more than 23 features. For example, consider the ``return type'' feature in the ``code analysis'' category. This can include numerous return types extracted from a given project, which could be void, int, URL, boolean, string, printStream, file, and date~(or a list of any of these periods).
Hence, as shown in Table \ref{table:distribution},
the number of features in our data varied from 39 to 60.

Note also that one way to summarize the results of this paper is that the golden set is an inaccurate, verbose description of the attributes required to detect static code warnings. 
% attributes required to defect static code attributes. 
As shown below, hiding within the 23 feature categories of Table~\ref{table:variables}, there exist two synthetic dimensions, which can be found via a linear SVM. 

\subsection{Evaluation Metrics}

Wang et al. reported their results in terms of 
{\em AUC} and {\em running time}:
\bi
\item
AUC~(Area Under the ROC Curve) measures the two-dimensional area under the Receiver Operator Characteristic (ROC) curve~\cite{witten2016data,heckman2011systematic}. It provides an aggregate and overall evaluation of performance across all possible classification thresholds to overall report the discrimination of a classifier~\cite{wang2018there}. This is a widely adopted measurement in the area of software engineering, especially for imbalanced data~\cite{liang2010automatic}. 
\item
Running time measures the efficiency of the execution of one algorithm. In this paper, we use the running time of one run from the start to the terminal of algorithm execution to compare the efficiency of different models.
\ei
Table~\ref{table:priorAUC} shows the AUC results achieved by Wang et al.~\cite{wang2018there}.   In summary, Wang et al. reported Random Forest as the best learner to identify actionable static warnings. 
% And there is much difference between Wang et al.'s and our results, which can be attributed to the different default parameters in Weka in Java~(tool employed by Wang et al.) and SciKit-Learn in Python~(tool employed in our paper).

In the software analytics literature, it is also common to assess learners via {\em recall} and {\em false alarms}:
\bi
\item
 Recall represents the ability of one algorithm to identify instances of positive class or actionable one from the given data set. It denotes the ratio of
detected actionable defects in comparison to the total number
of actionable defects in the data set generated by static warning tools, like FindBugs.
\item
 False Alarms
(pf) measures the instances or warnings generated from static warning tools falsely classified by an algorithm as positive or actionable which are actually negative or unactionable ones. This is an important index used to measure the efficiency of a defect prediction model.

\ei
In the following, we will report results for all of these four evaluation measures. 

\begin{table}[]
\footnotesize
\caption{\%AUC results reported in prior state-of-the-art~\cite{wang2018there} using proposed golden feature set.}
\begin{center}
\begin{tabular}{rccc}
 
\textbf{Project} & Random Forest & Decision Tree & SVM RBF \\ \hline
derby & 43 & 44 & 50 \\  
mvn & 45 & 45 & 50 \\  
lucence & 98 & 98 & 50 \\ 
phoenix & 71 & 70 & 62 \\  
cass & 70 & 69 & 67 \\  
jmeter & 86 & 82 & 50 \\  
tomcat & 80 & 64 & 50 \\  
ant & 44 & 44 & 50 \\  
commons & 57 & 56 & 50 \\ \hline
\textbf{median} & \textbf{70} & \textbf{64} & \textbf{50} 
\end{tabular}
\end{center}
\label{table:priorAUC}
\end{table}

%And the training process is done on Version 4 and testing on Version 5 of each project. 

%criteria on the 60 successive revisions with 3 or 6 months intervals. In this paper, we utilize Release 4 as training set and test on release 5.XXX need more details here

% not sure about getting rid of data imbalance, because we don't further explore this issue.

\subsection{Learning to Recognize Actionable Static Code Warnings}

Recall from the above that our data has two classes: actionable and non-actionable. Technically speaking, our task is a {\em binary classification problem}. A recent survey by Ghotra et al.~\cite{ghotra2015revisiting} found that for software analytics, the performance of dozens of binary classifications clusters into a handful of groups. Hence, by taking one classifier from each group, it is possible for just a few classifiers to act as representatives for a wide range of commonly used classifiers.

Decision trees~\cite{quinlan1987generating} seek splits to feature ranges that most minimize the diversity of classes  within each split. Once the best ``splitter'' is found, decision tree algorithms recurse on each split. 
% use skit learn tree.regressor or tree.classifier. if the former, reference CART brieman84 here. if the latter, reference Quinlan decision tres.

Random forests~\cite{breiman1999random} take the idea of decision trees one step further. Instead of building one tree, random forests build multiple trees (each time using a small random sample of the rows and columns from the original data). The final conclusion is then computed by a majority vote across all trees in the forest. 

Support vector machines~\cite{cortes1995support} take another approach. With a kernel function, the data is mapped into a higher-dimensional space. Then, using a quadratic programming, the algorithm finds the ``support vectors'' which are the instances closest to the boundary between to distinguish different classes.

\subsection{Deep Learning }
\label{tion:dl}

Since the Ghortra et al. survey~\cite{ghotra2015revisiting} was published in 2015, there has been much recent interest in the application of deep learning (DL) in software engineering. Applications of DL include bug localization~\cite{huo2019deep}, sentiment analysis~\cite{lin2018sentiment,guo2017semantically}, API mining~\cite{chen2019mining,gu2016deep,nguyen2017exploring}, effort estimation for agile development~\cite{choetkiertikul2018deep}, code similarity detection~\cite{zhao2018deepsim},
code clone detection~\cite{white2016deep}, etc.

% \begin{figure*}[!t]
%     \centering
%     {\includegraphics[width=\textwidth]{figs/cnn.png}}
%     \caption{Overview of CNN Model in Static Warning Identification.}
%     \label{fig:cnn}
% \end{figure*}

% \begin{figure*}[!t]
% \centerline
% {\includegraphics[width=0.5\textwidth,trim = {0 15cm 18cm 0}, clip]{figs/dnn1.jpeg}}
% \vspace*{-4cm}
% \caption{DLs comprised multiple layers of neurons.}   
% \label{fig:dnn}
% \end{figure*}

\begin{figure*}[!t]
\centerline
% \hspace*{-0.2cm}
%  trim={<left> <lower> <right> <upper>}
{\includegraphics[width=0.8\textwidth,trim = {0cm 6cm 40cm 0cm}, clip]{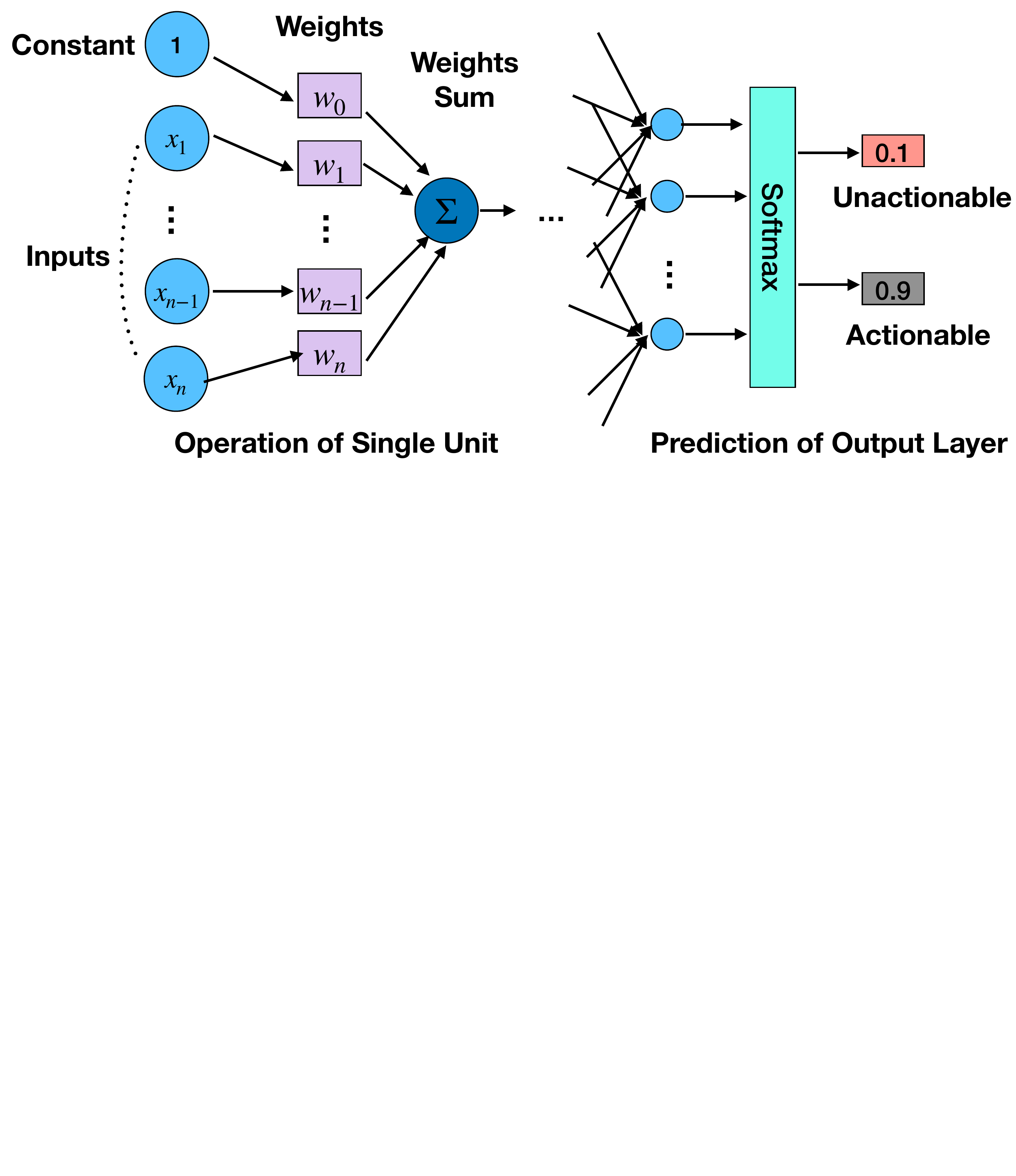}}
\vspace*{-6.5cm}
\caption{Illustration of DNN Model.}   
\label{fig:dnn}
\end{figure*}

Deep neural networks are layers of connected units called neurons.
% A brief architecture of five-layer fully connected DNN model is shown in Figure~\ref{fig:dnn}. 
A brief mechanism of fully connected DNN model is shown in Figure~\ref{fig:dnn}. 
For this paper, SE artifacts are transferred into vectors and fed into the neural networks as inputs in the input layer. Each neuron in hidden and output layers functions by multiplying its input with the weight of this neuron. Then the product is summed and then passed through a nonlinear transfer function called activation function to yield a variable. It either continuously serves as input to the next layer or final output of the network~\cite{goh1995back}.

Figure~\ref{fig:dnn} illustrates a layered architecture of neurons where inputs at layer $i$ are organized and synthesized as inputs at layer $i+1$ by non-linear transformations mentioned above. 
It's known as an automatic feature engineering model which efficiently extracts the non-linear and sophisticated patterns generally observed in the real world, like speech, video, audio.
For instance, technology-intensive companies like Google and Facebook are utilizing massive volumes of raw data for commercial data analysis ~\cite{najafabadi2015deep}. Within that layered architecture, only the most important signal from the inputs of layer $i$ will make it through to layer $i+1$. In this way, DL automates ``feature engineering'' which is the synthesis of important new features using some part or combination of other features. This, in turn, means that predictors can be learned from very complex input signals with multiple features, without requiring manual pre-processing. For example, Lin et al.~\cite{Lin2017StructuralDD} replaced their mostly manual analysis of $10^4$ features extracted from a wavelet package with a deep learner that automatically synthesized significant features.

DL trains its networks by running its data repeatedly through networks shown in Figure~\ref{fig:dnn} in multiple ``epochs''.
Each epoch pushes all the data by batch over the network and the resulting error on the output layer is computed. 
% This repeats until the average sum of squared error of training set is minimized.
This repeats until the training error or loss function on the validation set is minimized.
Error minimization is done via
back propagation~(BP). Parameters in DL (including neuron weights), are initialized randomly, and then these parameters of neurons are updated in each epoch of training using error back propagation.
Hornik et al.~\cite{hornik1991approximation} have shown that with sufficient hidden neurons, a single hidden layer back-propagation neural network can accurately approximate any continuous function.

% \begin{figure}[!t]
% \centerline{\includegraphics[width=0.8\textwidth]{dnn.jpeg}}
% \caption{\respto{2h1} \revised{Illustration of DNN Model. TODO: Embed a scalable image.}}   
% \label{fig:dnn}
% \end{figure}

DL training may require hundreds to thousands of epochs in complicated problems. However, overtraining makes the model overfit the training dataset and have poor generalization ability on the test set. 
Early stopping~\cite{zhang2016understanding} is a commonly used optimizer strategy and regulariser in deep learning, which improves generalization and prevents models from overfitting. It stops training when performance on a validation dataset starts to degrade. We tried to prevent overfitting in our domain via early stopping. The maximum epochs are set as 100, and the patience of early stopping as 3, i.e. stopping training DLs if the performance on the training set is not getting better for continuous three epochs.  After running our DLs, we could not improve performance after 8 to 30 epochs. Hence, all the results reported below come from 8 to 30 epochs.

\section{Experiments}
\label{sec:method}

\subsection{Learning Schemes}

For this study, the non-DL learners came from SciKit-Learn~\cite{pedregosa2011scikit} while the DL methods came from the Keras package~\cite{geron2019hands}. For the three non-DL learners (Random Forests, Decision Tree, linear Support vector machines), we ran these using their default control settings from SciKit-Learn.
As to Deep Learning, we ran three DL schemes. As suggested in the literature review~\cite{li2018deep}, (fully-connected) deep neural network~(DNN) and convolutional neural network~(CNN) are mostly explored DL models in SE area.

The first scheme is a fully connected deep neural network~(DNN).
For a description of this method, see Section~\tion{dl}.
Starting with the defaults from Keras, we configure our DNN model as follows:
\bi
\item
5 fully connected layers (with 30 neurons for each hidden layer) concatenated by dropout layers in between. 
\item
The activation functions for hidden layers were implemented using the {\em Relu} function. Relu represents a rectified linear unit, whose formula is denoted as $ f(x) = max(0, x) $. As a universal choice of various activation functions, Relu is known for many merits like fast to compute and converge in practice and its gradients not vanishing when $ x > 1$ holds or the current neuron is activated~\cite{li2017convergence}. Batch normalization layers are conducted before each activation function to avoid the internal covariate shift~(with the distribution changes of parameters in training deep neural networks, the current layer has to constantly readjust to new distributions)~\cite{ioffe2015batch}.
\item
As said above, actionable warning identification is a binary problem. That is, for any instance $ i $ of warnings, its label $ y_i \in\{0, 1\} $, where $ 0 $ denotes this warning is unactionable and $ 1 $ denotes as actionable. Consequently, we use softmax as the activation function for the output of our network in the output layer. Softmax takes the vectors generated from the last hidden layer as inputs and proceeds them by exponentiation operation with a power of $e$ and mapping it into a list of probability distribution of all the label class candidates. For each instance, the list of Softmax vector $ [P_0, P_1] $ generated from softmax function always sums to 1, where $ P_0$ is the probability that this bug is unactionable while $ P_1$ denoted as actionable.
\ei
% Softmax also works for multi-label classification problem, the formula is defined as follows:
% \[ P_i =
% Softmax(y_i)
% \frac{e^{y_i}}{\sum_{j=1}^N e^{y_i}}   \]
% \[where: \sum_{i=1}^N P_i = 1\]

%  trim={<left> <lower> <right> <upper>}
\begin{figure*}
% \begin{figure*}[!t]
\centerline
% \hspace*{-0.2cm}
{\includegraphics[width=1.1\textwidth,trim = {0 8cm 9.5cm 0}, clip]{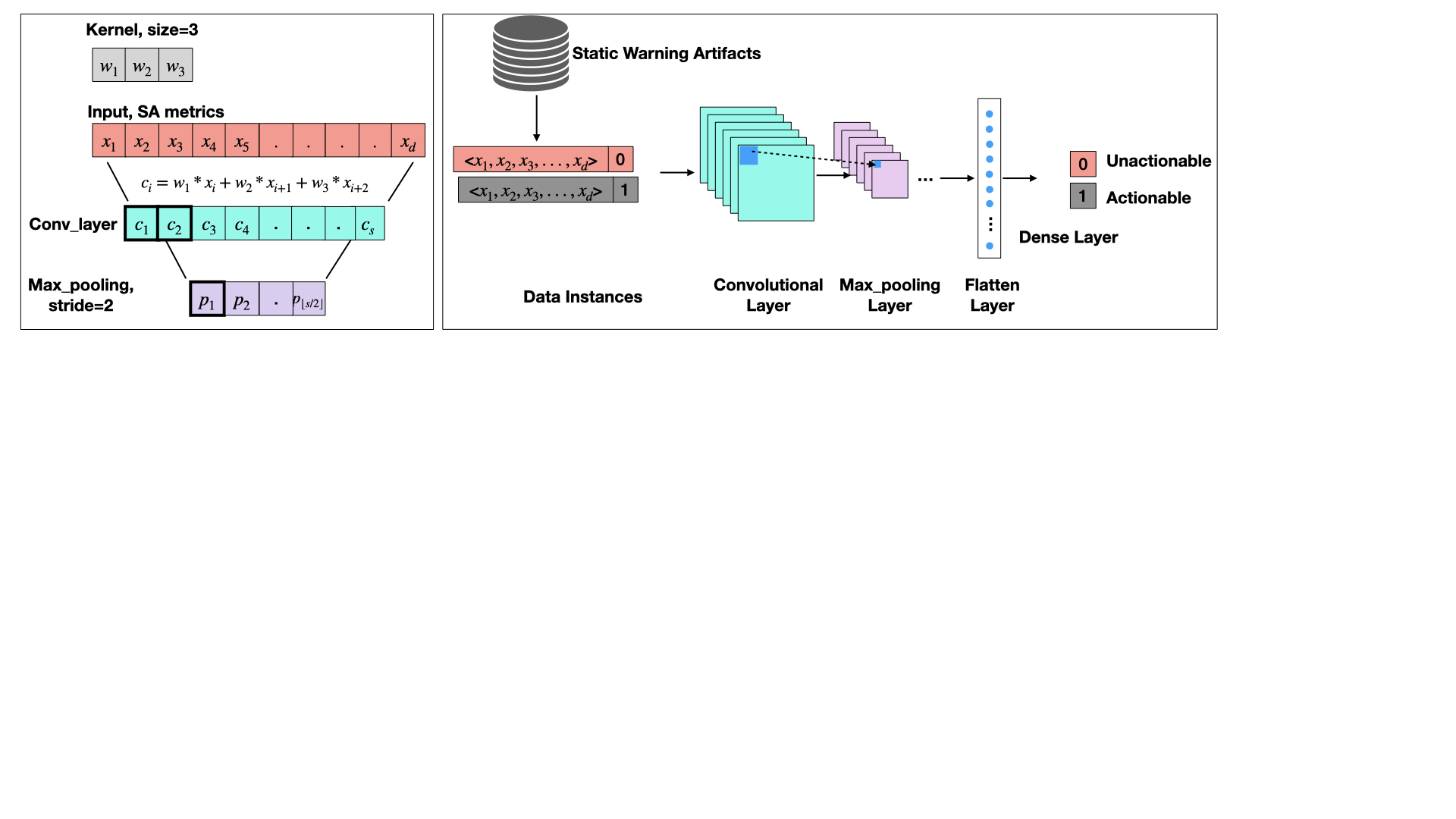}}
\vspace*{-3.5cm}
\caption{Overview of CNN Model in Static Warning Identification.}   
\label{fig:cnn}
\end{figure*}

Our second scheme is CNN (convolutional neural network)~\cite{goodfellow2016deep}, a widely used DL method which employs weight sharing and pooling schemes. Figure~\ref{fig:cnn} illustrates the overview scheme of applying CNN in static warning analysis. Convolutional layers work with a filter of inputs to build a feature map for repeated times, whose principle is looking for correlation between filter and input feature matrix. And max pooling layers reduce spatial size of features by selecting maximum value to represent a feature window. With weight sharing of filters and max pooling, CNNs can greatly reduces the parameters required in training phase. 

% For example, in the ImageNet competition in 2012, CNN achieved competitive results to a state-of-the-art approach while at the same time halving the error rate~\cite{krizhevsky2012imagenet}.

DNN\_weighted is our third DL scheme whose main structure is the same as DNN mentioned above but also use a weighted strategy. 
Table~\ref{table:distribution} shows that many of our data sets have {\em unbalanced class distributions}
where our target class (actionable warnings) is very under-represented  (often less than 20\%).
To address this data imbalance problem, we re-weight the minority class, actionable class. Specifically, we use the reciprocal of the ratio for class 0 and 1 to weight the loss function during the training phase. For instance, the ratio of actionable samples in training set is 0.25, the weighting scheme sets the weight of actionable (minority) as 4, and unactionable (majority) as 1 to balance the significance of training loss for two classes in the training process.
Note that we used this reweighting scheme rather than some alternative method (e.g. duplicate instances of minority class) since reusing many copies of one instance in the training set causes extra computational cost~\cite{shalev2014understanding}.

\subsection{Statistical Tests}\label{tion:stats}

To select ``best'' learning methods, the advice of Rosenthal et al.~\cite{rosenthal1994parametric} is taken in this paper. Specifically, given that all our numbers are with 0..1, then experiment results are not prone to extreme outlier effects via statistical tests. Such extreme outliers are indicators for long-tail effects which, in turn suggest that it might be better to use non-parametric methods. This is not ideal since non-parametric tests have less statistical power than parametric ones. 

Rosenthal et al. discuss different parametric methods for asserting that one result is with some small effect of another (i.e. it is ``close to'').
They list dozens of effect size tests that divide into two groups: the $r$ group that is based on the Pearson correlation coefficient; or the $d$ family that is based on absolute differences normalized by (e.g.) the size of the standard deviation. Rosenthal et al. comment that ``none is intrinsically better than the other''. The most direct method is utilized in our paper. Using a $d$ family method, it can be concluded that one distribution is the same as another if their mean value differs by less than Cohen's delta ($d$*standard deviation). 
Note that $d$ is computed  separately for each different evaluation measure~(recall, false alarm, AUC).
 
To visualize that ``close to'' analysis, in all our results:
\bi
\item
Any cell that is within $d$ of the best value will be highlighted in gray.
All gray cells are observed as ``winners'' and all the other cells are ``losers''.
\item
For recall and AUC, the ``best'' cells have ``highest value'' since the optimization goal is to maximize these values.
\item
For false alarm, the ``best'' cells have ``lowest value'' since false alarms is to be minimized. 
\ei

As to what value of $d$ to use in this analysis, we take the advice of
a widely cited paper by
Sawilowsky~\cite{sawilowsky2009new}
(this 2009 paper has 1100 citations). That paper
asserts that ``small'' and ``medium'' effects can be measured using $d=0.2$ and $d=0.5$ (respectively).
Splitting the difference, we will analyze this data looking for differences larger than $d=(0.5+0.2)/2=0.35$.

\subsection{Results}
\label{sec:experiment}

In the text of {\em Empirical AI}, Cohen advises that any method uses a random number generator must be run multiple times, to allow for any effects introduced by the random number seed. For deterministic models, the same output is always produced for the same sequence of given a particular input. To dispel the bias between deterministic and non-deterministic models and eliminate the bias of uncertainty: 

 \bi
 \item
 Ten times, 
 we shuffled the training and test data into some random order.
 \item
 Each time, divide the test data into five bins, taking care to implement {\em stratified sampling}; i.e. ensuring that the class distribution of the whole data is replicated within each bin. In this way, the distributions of training and testing set are kept unchanged, as shown in Table~\ref{table:distribution}.

% give examples like PCA
\item For each 20\% test bins, learn a model using 100\% of the training set.
 
\ei

\begin{table*}[!t]
\caption{Summary results of recall, false alarm and AUC on nine datasets. Cells in gray denote the ``best'' results for each row, where ``best'' means within  $d$ difference to the best value (and $d$ is calculated as per \tion{stats}.)} 
\large
\begin{adjustbox}{max width=\textwidth}
\renewcommand{\arraystretch}{1.5}
\begin{tabular}{|l|r|rrrrrr|}
\hline
\textbf{}                            & \multicolumn{1}{l}{\cellcolor[HTML]{020202}{\color[HTML]{FFFFFF} \textbf{Project}}} & \multicolumn{1}{l}{\cellcolor[HTML]{020202}{\color[HTML]{FFFFFF} \textbf{DNN weighted}}} & \multicolumn{1}{l}{\cellcolor[HTML]{020202}{\color[HTML]{FFFFFF} \textbf{CNN}}} & \multicolumn{1}{l}{\cellcolor[HTML]{020202}{\color[HTML]{FFFFFF} \textbf{DNN}}} & \multicolumn{1}{l}{\cellcolor[HTML]{020202}{\color[HTML]{FFFFFF} \textbf{Random Forest}}} & \multicolumn{1}{l}{\cellcolor[HTML]{020202}{\color[HTML]{FFFFFF} \textbf{Decision Tree}}} & \multicolumn{1}{l}{\cellcolor[HTML]{020202}{\color[HTML]{FFFFFF} \textbf{SVM linear}}} \\\hline
\textbf{}  & \textbf{derby}  & \cellcolor[HTML]{C0C0C0}\textbf{96.6\%}  &\cellcolor[HTML]{C0C0C0}\textbf{96.9\%} & \textbf{94.0\%} & \textbf{92.0\%} & \cellcolor[HTML]{C0C0C0}\textbf{94.8\%} & \cellcolor[HTML]{C0C0C0}\textbf{97.8\%} \\
\textbf{}  & \textbf{mvn}  & \cellcolor[HTML]{C0C0C0}\textbf{97.6\%} & \cellcolor[HTML]{C0C0C0}\textbf{95.0\%}& \textbf{92.0\%}  & \textbf{78.9\%}  & \cellcolor[HTML]{C0C0C0}\textbf{94.7\%} & \cellcolor[HTML]{C0C0C0}\textbf{97.0\%}  \\
\textbf{} & \textbf{lucence}  & \cellcolor[HTML]{C0C0C0}\textbf{95.3\%} & \cellcolor[HTML]{C0C0C0}\textbf{98.1\%}  & \textbf{91.3\%}  & \cellcolor[HTML]{C0C0C0}\textbf{96.8\%} & \cellcolor[HTML]{C0C0C0}\textbf{96.6\%}  & \textbf{87.1\%}  \\
\textbf{Recall}  & \textbf{phoenix} & \cellcolor[HTML]{C0C0C0}\textbf{95.2\%} & \textbf{93.0\%}   & \textbf{89.3\%}  & \textbf{88.7\%}  & \textbf{86.8\%} & \cellcolor[HTML]{C0C0C0}\textbf{96.1\%}  \\
\textbf{}  & \textbf{cass}  & \textbf{81.3\%} & \cellcolor[HTML]{C0C0C0}\textbf{98.8\%}  & \textbf{68.1\%} & \textbf{76.8\%} & \textbf{75.7\%}  & \textbf{90.3\%} \\
\textbf{(d = 3\%)}  & \textbf{jmeter}   & \cellcolor[HTML]{C0C0C0}\textbf{94.3\%}  & \cellcolor[HTML]{C0C0C0}\textbf{93.9\%}  & \textbf{89.2\%}   & \cellcolor[HTML]{C0C0C0}\textbf{96.9\%}  & \textbf{92.7\%}  & \textbf{93.3\%}  \\
\textbf{}   & \textbf{tomcat}   & \cellcolor[HTML]{C0C0C0}\textbf{98.0\%}  & \textbf{95.0\%} & \cellcolor[HTML]{C0C0C0}\textbf{96.4\%} & \textbf{91.8\%}  & \textbf{87.6\%}  & \cellcolor[HTML]{C0C0C0}\textbf{98.2\%}  \\
& \textbf{ant} & \textbf{91.1\%}  & \cellcolor[HTML]{C0C0C0}\textbf{93.1\%}  & \textbf{84.1\%} & \textbf{78.7\%}  & \textbf{87.0\%} & \cellcolor[HTML]{C0C0C0}\textbf{95.0\%} \\
\textbf{}  & \textbf{commons}  & \textbf{81.1\%}  & \cellcolor[HTML]{C0C0C0}\textbf{97.8\%} & \textbf{73.3\%}  & \textbf{66.7\%}   & \textbf{92.0\%}   & \cellcolor[HTML]{C0C0C0}\textbf{99.5\%} \\\hline
\textbf{}  & \textbf{derby}  & \cellcolor[HTML]{C0C0C0}\textbf{1.2\%} & \textbf{10.8\%} & \cellcolor[HTML]{C0C0C0}\textbf{0.5\%}  & \cellcolor[HTML]{C0C0C0}\textbf{0.3\%}   & \cellcolor[HTML]{C0C0C0}\textbf{0.5\%} & \cellcolor[HTML]{C0C0C0}\textbf{1.3\%} \\  & \textbf{mvn}  & \cellcolor[HTML]{C0C0C0}\textbf{1.4\%}  & \textbf{6.8\%}  & \cellcolor[HTML]{C0C0C0}\textbf{0.4\%}  & \cellcolor[HTML]{C0C0C0}\textbf{0.5\%} & \cellcolor[HTML]{C0C0C0}\textbf{0.4\%} & \cellcolor[HTML]{C0C0C0}\textbf{1.2\%} \\
\textbf{}  & \textbf{lucence}   & \textbf{5.9\%}  & \textbf{5.8\%}   & \cellcolor[HTML]{C0C0C0}\textbf{3.2\%}   & \cellcolor[HTML]{C0C0C0}\textbf{1.4\%}    & \cellcolor[HTML]{C0C0C0}\textbf{3.2\%}   & \textbf{6.9\%}  \\
\multicolumn{1}{|c|}{\textbf{False}} & \textbf{phoenix}   & \textbf{3.0\%}  & \textbf{8.7\%}  & \cellcolor[HTML]{C0C0C0}\textbf{1.4\%}   & \cellcolor[HTML]{C0C0C0}\textbf{1.3\%}   & \cellcolor[HTML]{C0C0C0}\textbf{0.7\%}  & \textbf{3.5\%}  \\
\multicolumn{1}{|c|}{\textbf{Alarm}}   & \textbf{cass}    & \cellcolor[HTML]{C0C0C0}\textbf{1.2\%}  & \textbf{7.0\%}   & \cellcolor[HTML]{C0C0C0}\textbf{0.4\%}  & \textbf{2.5\%}    & \cellcolor[HTML]{C0C0C0}\textbf{1.3\%}  & \cellcolor[HTML]{C0C0C0}\textbf{1.4\%}  \\ 
\textbf{}  & \textbf{jmeter}   & \textbf{3.1\%}  & \textbf{48.6\%}   & \cellcolor[HTML]{C0C0C0}\textbf{1.4\%} & \cellcolor[HTML]{C0C0C0}\textbf{1.3\%}    & \cellcolor[HTML]{C0C0C0}\textbf{0.4\%}   & \cellcolor[HTML]{C0C0C0}\textbf{2.1\%}   \\
\textbf{} & \textbf{tomcat}   & \cellcolor[HTML]{C0C0C0}\textbf{2.1\%}   & \textbf{8.8\%}  & \cellcolor[HTML]{C0C0C0}\textbf{1.3\%} & \cellcolor[HTML]{C0C0C0}\textbf{0.4\%} & \textbf{4.3\%} & \textbf{3.2\%} \\
\textbf{(d=2\%)} & \textbf{ant} & \cellcolor[HTML]{C0C0C0}\textbf{0.5\%} & \textbf{6.7\%}  & \cellcolor[HTML]{C0C0C0}\textbf{0.5\%}& \cellcolor[HTML]{C0C0C0}\textbf{0.4\%} & \cellcolor[HTML]{C0C0C0}\textbf{0.5\%} & \cellcolor[HTML]{C0C0C0}\textbf{0.5\%}\\
\textbf{} & \textbf{commons} & \textbf{3.1\%} & \textbf{8.6\%} & \cellcolor[HTML]{C0C0C0}\textbf{1.4\%} & \cellcolor[HTML]{C0C0C0}\textbf{0.2\%} & \cellcolor[HTML]{C0C0C0}\textbf{1.4\%} & \textbf{5.8\%} \\\hline
\textbf{} & \textbf{derby}   & \cellcolor[HTML]{C0C0C0}\textbf{99.7\%} & \textbf{97.2\%} & \cellcolor[HTML]{C0C0C0}\textbf{99.6\%}& \cellcolor[HTML]{C0C0C0}\textbf{99.7\%} & \textbf{97.1\%}  & \cellcolor[HTML]{C0C0C0}\textbf{99.5\%} \\
\textbf{}  & \textbf{mvn}  & \cellcolor[HTML]{C0C0C0}\textbf{99.9\%} & \cellcolor[HTML]{C0C0C0}\textbf{99.1\%} & \cellcolor[HTML]{C0C0C0}\textbf{99.9\%} & \cellcolor[HTML]{C0C0C0}\textbf{99.6\%} & \textbf{96.8\%} & \cellcolor[HTML]{C0C0C0}\textbf{99.6\%} \\
\textbf{}  & \textbf{lucence} & \cellcolor[HTML]{C0C0C0}\textbf{98.7\%} & \cellcolor[HTML]{C0C0C0}\textbf{98.7\%}  & \cellcolor[HTML]{C0C0C0}\textbf{98.8\%} & \cellcolor[HTML]{C0C0C0}\textbf{99.6\%} & \textbf{96.6\%} & \textbf{97.3\%} \\
\multicolumn{1}{|c|}{\textbf{AUC}}  & \textbf{phoenix}  & \cellcolor[HTML]{C0C0C0}\textbf{98.5\%} & \cellcolor[HTML]{C0C0C0}\textbf{97.8\%}  & \cellcolor[HTML]{C0C0C0}\textbf{98.8\%}  & \cellcolor[HTML]{C0C0C0}\textbf{98.6\%}  & \textbf{92.7\%}  & \cellcolor[HTML]{C0C0C0}\textbf{98.8\%}\\
\textbf{}  & \textbf{cass}  & \textbf{97.0\%} & \textbf{98.0\%}  & \textbf{96.9\%} & \textbf{98.6\%} & \textbf{88.0\%} & \cellcolor[HTML]{C0C0C0}\textbf{99.7\%} \\
\multicolumn{1}{|c|}{\textbf{(d=1\%)}} & \textbf{jmeter} & \cellcolor[HTML]{C0C0C0}\textbf{98.7\%} & \textbf{82.3\%}  & \textbf{97.7\%}  & \cellcolor[HTML]{C0C0C0}\textbf{99.7\%}  & \textbf{95.9\%} & \cellcolor[HTML]{C0C0C0}\textbf{98.8\%} \\
\textbf{} & \textbf{tomcat}  & \cellcolor[HTML]{C0C0C0}\textbf{100.0\%} & \textbf{98.3\%}  & \cellcolor[HTML]{C0C0C0}\textbf{99.7\%} & \cellcolor[HTML]{C0C0C0}\textbf{99.6\%}  & \cellcolor[HTML]{FFFFFF}\textbf{92.1\%} & \cellcolor[HTML]{C0C0C0}\textbf{99.6\%}  \\
\textbf{}  & \textbf{ant}  & \cellcolor[HTML]{C0C0C0}\textbf{98.9\%} & \textbf{97.3\%} & \textbf{97.7\%} & \cellcolor[HTML]{C0C0C0}\textbf{98.7\%} & \textbf{93.3\%} & \cellcolor[HTML]{C0C0C0}\textbf{99.7\%} \\
\textbf{} & \textbf{commons} & \textbf{96.0\%} & \cellcolor[HTML]{C0C0C0}\textbf{99.2\%} & \cellcolor[HTML]{FFFFFF}\textbf{97.7\%} & \cellcolor[HTML]{C0C0C0}\textbf{98.7\%} & \cellcolor[HTML]{FFFFFF}\textbf{96.1\%}  & \cellcolor[HTML]{C0C0C0}\textbf{99.0\%}              \\\hline                 
\end{tabular}
\label{tbl:summary}
\end{adjustbox}
\end{table*}

% \begin{table*}[!t]

% \caption{\respto{2h2}\revised{TODO: typeset as a table.} Summary results of recall, false alarm and AUC on nine datasets. Cells in gray denote the ``best'' results for each row, where ``best'' means within  $d$ difference to the best value (and $d$ is calculated as per \tion{stats}.)}
% % \caption{Summery for median results on ten repeated runs (recall, false alarm and AUC on nine datasets). Cells in gray denote results  the ``best'' results for each row, where ``best'' means within  $d$ difference to the best value (and $d$ is calculated as per \tion{stats}.)}
% \label{tbl:summary}
% \begin{center}
% \includegraphics[width=4.5in]{fig_100vs_20d.png}
% \end{center}
% \end{table*}

% Please add the following required packages to your document preamble:
% \usepackage[table,xcdraw]{xcolor}
% If you use beamer only pass "xcolor=table" option, i.e. \documentclass[xcolor=table]{beamer}
\begin{table}[!t]
\caption{Comparing median results and IQR of recall, false alarm and AUC. Cells in gray denote the ``best'' median results for each row, where ``best'' means within  $d$ difference to the best value  in each row (and $d$ is calculated as per \tion{stats}.)} 
\large
\begin{adjustbox}{max width=\textwidth}
\renewcommand{\arraystretch}{1.5}
\begin{tabular}{|l|lrrrrrr|}
\rowcolor[HTML]{01000C} 
{\color[HTML]{FFFFFF} \textbf{metrics}} & \cellcolor[HTML]{020202}{\color[HTML]{FFFFFF} \textbf{measures}} & {\color[HTML]{FFFFFF} \textbf{DNN weighted}} & {\color[HTML]{FFFFFF} \textbf{CNN}} & {\color[HTML]{FFFFFF} \textbf{DNN}} & {\color[HTML]{FFFFFF} \textbf{Random Forest}} & {\color[HTML]{FFFFFF} \textbf{Decision Tree}} & {\color[HTML]{FFFFFF} \textbf{SVM linear}} \\
\textbf{recall} & \textbf{median} & \cellcolor[HTML]{C0C0C0}\textbf{95.2\%}  & \cellcolor[HTML]{C0C0C0}\textbf{95.1\%}  & \textbf{89.3\%} & \textbf{88.7\%}  & \textbf{92.0\%}  & \cellcolor[HTML]{C0C0C0}\textbf{96.1\%}  \\
\textbf{(d=1\%)} & \textbf{IQR}  & \textbf{5.4\%}   & \textbf{3.9\%}  & \textbf{8.0\%}  & \textbf{13.3\%}  & \textbf{7.7\%}  & \textbf{4.5\%}  \\\hline
\textbf{false alarm}  & \textbf{median}  & \textbf{2.1\%}  & \textbf{8.6\%}  & \cellcolor[HTML]{C0C0C0}\textbf{1.3\%} & \cellcolor[HTML]{C0C0C0}\textbf{0.5\%} & \cellcolor[HTML]{C0C0C0}\textbf{0.7\%} & \textbf{2.1\%} \\
\textbf{(d=1\%)} & \textbf{IQR} & \textbf{1.9\%} & \textbf{2.0\%}  & \textbf{0.9\%} & \textbf{0.9\%}  & \textbf{0.9\%}  & \textbf{2.2\%} \\\hline
\textbf{AUC}  & \textbf{median}   & \textbf{98.7\%}  & \textbf{98.0\%}  & \textbf{98.8\%}  & \cellcolor[HTML]{C0C0C0}\textbf{99.5\%}  & \textbf{95.9\%}  & \cellcolor[HTML]{C0C0C0}\textbf{99.5\%}  \\
\textbf{(d=0\%)}  & \textbf{IQR}  & \textbf{1.2\%} & \textbf{1.5\%} & \textbf{1.9\%} & \textbf{0.9\%}  & \textbf{3.9\%} & \textbf{0.8\%}   \\\hline         \end{tabular}
\end{adjustbox}
\label{tbl:median}
\end{table}

% \begin{table*}[!t]
% \caption{\respto{2h3}\revised{TODO: typeset as a table.} Comparing median results and IQR of recall, false alarm and AUC. Cells in gray denote the ``best'' median results for each row, where ``best'' means within  $d$ difference to the best value  in each row (and $d$ is calculated as per \tion{stats}.)}
% \begin{center}
% \includegraphics[width=4.5in]{summery_median_100vs20d.png}
% \end{center}
% \label{tbl:median}
% \end{table*}

\renewcommand{\bfdefault}{bx}
% Please add the following required packages to your document preamble:
% \usepackage[table,xcdraw]{xcolor}
% If you use beamer only pass "xcolor=table" option, i.e. \documentclass[xcolor=table]{beamer}
\begin{table*}[!t]
\caption{Comparing results of running time sorted by size of datasets in a descending order on nine projects from six learners.} 
\large
\begin{adjustbox}{max width=\textwidth}
\renewcommand{\arraystretch}{1.5}
\begin{tabular}{|l|r|rrrrrr|} \hline
\textbf{}            & \multicolumn{1}{l}{\cellcolor[HTML]{020202}{\color[HTML]{FFFFFF} \textbf{Project}}} & \multicolumn{1}{l}{\cellcolor[HTML]{020202}{\color[HTML]{FFFFFF} \textbf{DNN weighted}}} & \multicolumn{1}{l}{\cellcolor[HTML]{020202}{\color[HTML]{FFFFFF} \textbf{CNN}}} & \multicolumn{1}{l}{\cellcolor[HTML]{020202}{\color[HTML]{FFFFFF} \textbf{DNN}}} & \multicolumn{1}{l}{\cellcolor[HTML]{020202}{\color[HTML]{FFFFFF} \textbf{Random Forest}}} & \multicolumn{1}{l}{\cellcolor[HTML]{020202}{\color[HTML]{FFFFFF} \textbf{Decision Tree}}} & \multicolumn{1}{l}{\cellcolor[HTML]{020202}{\color[HTML]{FFFFFF} \textbf{SVM linear}}} \\\hline
\textbf{}  & \textbf{lucence}  & \textbf{188.3} & \textbf{233.1}  & \textbf{201.0} & \textbf{3.9} & \textbf{3.2}& \textbf{25.8}\\
\textbf{}  & \textbf{cass} & \textbf{172.5}  & \textbf{368.1} & \textbf{195.6} & \textbf{2.4}  & \textbf{2.5} & \textbf{7.9} \\
\textbf{}  & \textbf{derby}  & \textbf{145.0} & \textbf{351.7} & \textbf{156.5} & \textbf{2.7} & \textbf{2.3} & \textbf{7.7} \\
\textbf{} & \textbf{phoenix}  & \textbf{134.6}  & \textbf{293.6}  & \textbf{148.2} & \textbf{2.1}  & \textbf{2.1}   & \textbf{6.1}  \\
\textbf{Runtime(/s)} & \textbf{tomcat} & \textbf{119.2}  & \textbf{280.6}  & \textbf{112.5}  & \textbf{1.8} & \textbf{1.4} & \textbf{3.4} \\
\textbf{} & \textbf{ant} & \textbf{110.6} & \textbf{259.4} & \textbf{123.5} & {\color[HTML]{333333} \textbf{1.8}}  & \textbf{1.3} & \textbf{2.4} \\
\textbf{}  & \textbf{mvn}  & \textbf{70.4} & \textbf{114.6}  & \textbf{73.7} & \textbf{1.9} & \textbf{0.9} & \textbf{1.6} \\
\textbf{}  & \textbf{commons} & \textbf{86.5} & \textbf{124.5} & \textbf{96.1} & \textbf{1.5} & \textbf{0.7} & \textbf{1.1}  \\
\textbf{}  & \textbf{jmeter}  & \textbf{57.0}  & \textbf{110.8}  & \textbf{60.8} & \textbf{1.5}  & \textbf{0.8} & \textbf{1.3} \\ \cline{2-8}
\textbf{}   & \textbf{Average}  & \textbf{120.5} & \textbf{237.4}   & \textbf{129.8}  & \textbf{2.2}   & \textbf{1.7}  & \textbf{6.4}     \\\hline                          
\end{tabular}
\end{adjustbox}
\label{tbl:runtime}
\end{table*}

% \begin{table*}[!t]
% \begin{center}
% \includegraphics[width=4.5in]{runtime_100vs20_1.png}
% \end{center}
% \caption{\respto{2h4}\revised{TODO: typeset as a table.} Comparing results of running time sorted by size of datasets in a descending order on nine projects from six learners.}
% \label{tbl:runtime}
% \end{table*}

\tbl{summary} shows the results of our experiment rig.
The gray cells
show results
that are either
(a)~the best values or (b)~are as good as the best. 
Counting  the winning gray cells and the other white cells,
we can see that:
\bi
\item
Linear SVM are often  preferred  (lower false alarms, higher recall and AUC).
\item
The tree learners  have many white cells; i.e. they perform worse than   best.
\item
The deep learners (DNN weighted, CNN, DNN) are often gray-- but not as often as SVM linear.
\ei
Hence we say
that linear SVM has the best all-around performance.

Another reason to prefer SVMs over deep learners is shown in 
\tbl{runtime}.
This table
shows the runtimes of our different learners:
deep learners were very much slower than the other learners (at least 20 times faster). 
% where 100\% of the data from release $i$ was used for training and shuffled and stratified 20\% of data from release $i+1$ was used for testing. Five times test results of recall, false alarm and AUC on 5 bins of testset are averaged respectively. To eliminate the bias of uncertainty, average performance of ten repeated runs is reported finally. 

Note that, compared with Table~\ref{table:priorAUC}, our AUC results shown in Table~\ref{tbl:summary} and Table~\ref{tbl:median} are much better than Wang et al.'s,
which we explain as follows.
Firstly, the default parameters in Weka (used    by Wang et al.) are different to those used in SciKit-Learn ~(the tool employed in our paper). 

Secondly, we use a different SVM to Wang et al.
In \tbl{summary}, Random Forest performs best in baseline models from the perspective of AUC which is consistent with Wang et al. While SVM result indicates significant difference due to different choices of kernels. (We also conducted an experiment on SVM with RBF kernel and got median AUC as 0.5.)

In summary, we can endorse the use of linear SVM in this domain, but not deep learners or tree learners.

\section{Why Such Similar Performance?}
\label{sec:intrinsic}
 
A question raised by the above results is 
why do different learners perform so similarly on 
all these data sets. Accordingly, this section explores that issue.

We will argue that the above results illustrates 
Vandekerckhove et al. {\em Principle of Parsimony}.
They 
warn that unnecessary sophisticated models 
% not only burdens computational cost, 
can damage the generalization capability of the classifiers~\cite{vandekerckhove2015model}. 
This
principle is a strategy that warns against overfitting
(and  is a fundamental principle of model selection). It 
suggests
that simpler models are preferred than complex ones if those models obtain similar performance.

A convincing demonstration that Principle of Parsimony has two
parts:
  \be
  \item We must show some {\em  
  damage to the generalization capability of a complex classifier}.
  For example, in the above, we found that
  even though
  deep learner's automatic feature engineering
  may account for irrelevant particulars (like noise in the data),
 they did not perform better than
  linear SVM.
%   \item We are using a learner that is  {\em using many dimensions}. In our case, this is easy to demonstrate. 
%  Recalling \tion{dl}, deep learners are automated feature engineering tools that routinely combine together many influences in order to achieve some result.
 \item We must also show that the  data set  has only
{\em very few dimensions}; i.e. a complex learner  is exploring
an inherently simple set of data. 
In the rest of this section, 
using an {\em intrinsic dimensionality calculator},
we will show that the intrinsic dimensionality of our static warning data sets is never more than two and usually is less.
\ee
To say all that another way, since the problem explored in our study is inherently low dimensional,
it is hardly surprising that the sophistication of deep learning was not useful in this domain.

\subsection{What is ``Intrinsic Dimensionality''?}\label{tion:it}
  Levina et al.~\cite{levina2005maximum} comment that the reason any data mining method works for high dimensions is that data embedded in high-dimensional format actually can be converted into a more compressed space without major
  information loss. 
A traditional way to compute these
intrinsic dimensions is PCA (Principal Component Analysis).
But Levina et al. caution that, as data in the real-world becomes increasingly sophisticated and non-linearly decomposable, PCA methods tend to overestimate the dimensions of a  data set~\cite{levina2005maximum}.

% \textcolor{MidnightBlue}
Instead, Levina et al. propose a fractal-based method for calculating
intrinsic dimensionality
(and that method is now a standard technique in other fields such as astrophysics).
 The intrinsic dimension of a data set with \textit{N}
items is found by computing the number of items found at distance within radius \textit{r} (where \textit{r} is the  distance between two configurations) while varying \textit{r}.
This measures the intrinsic dimensionality since:
\bi
\item If the items spread out in only one $r=1$ dimensions, then we will only find linearly more items as $r$ increases.
\item But the items spread out in, say, $r>1$ dimensions, then we will find polynomially more items as $r$ increases.
\ei

As shown in Equation~\ref{eq:cr},
Levina et al. normalize the number of items found according to the number of $N$ items being compared.
They recommend reporting the number of intrinsic dimensions as the maximum value of the slope between  $\mathit{ln(r)}$ vs  the $\mathit{ln}(C(r))$ value computed as follows.
Note Equation~\ref{eq:cr} use the L1-norm to calculate distance rather than the  Euclidean L2-norm. 
As seen in Table~\ref{table:distribution}, our raw data has up to 60 dimensions.
Courtney et al.~\cite{Aggarwal01} advise that for such high dimensional data, L1 performs better than L2.

\begin{equation}\label{eq:cr}
C(r) = \frac{2}{N(N-1)} \sum_{i=1}^N \sum_{j=i + 1}^N I (||x_i,x_j||<r) 
\end{equation}
\[where: I (||x_i,x_j||<r) = \left\{
    \begin{aligned}
    1, ||x_i,x_j|| < r \\
    0, ||x_i,x_j|| \ge r
    \end{aligned}
    \right. \]
    \noindent
    
\begin{figure} 
\centerline{\includegraphics[width=0.45\textwidth,trim = {0 0 1.3cm 0.8cm}, clip]{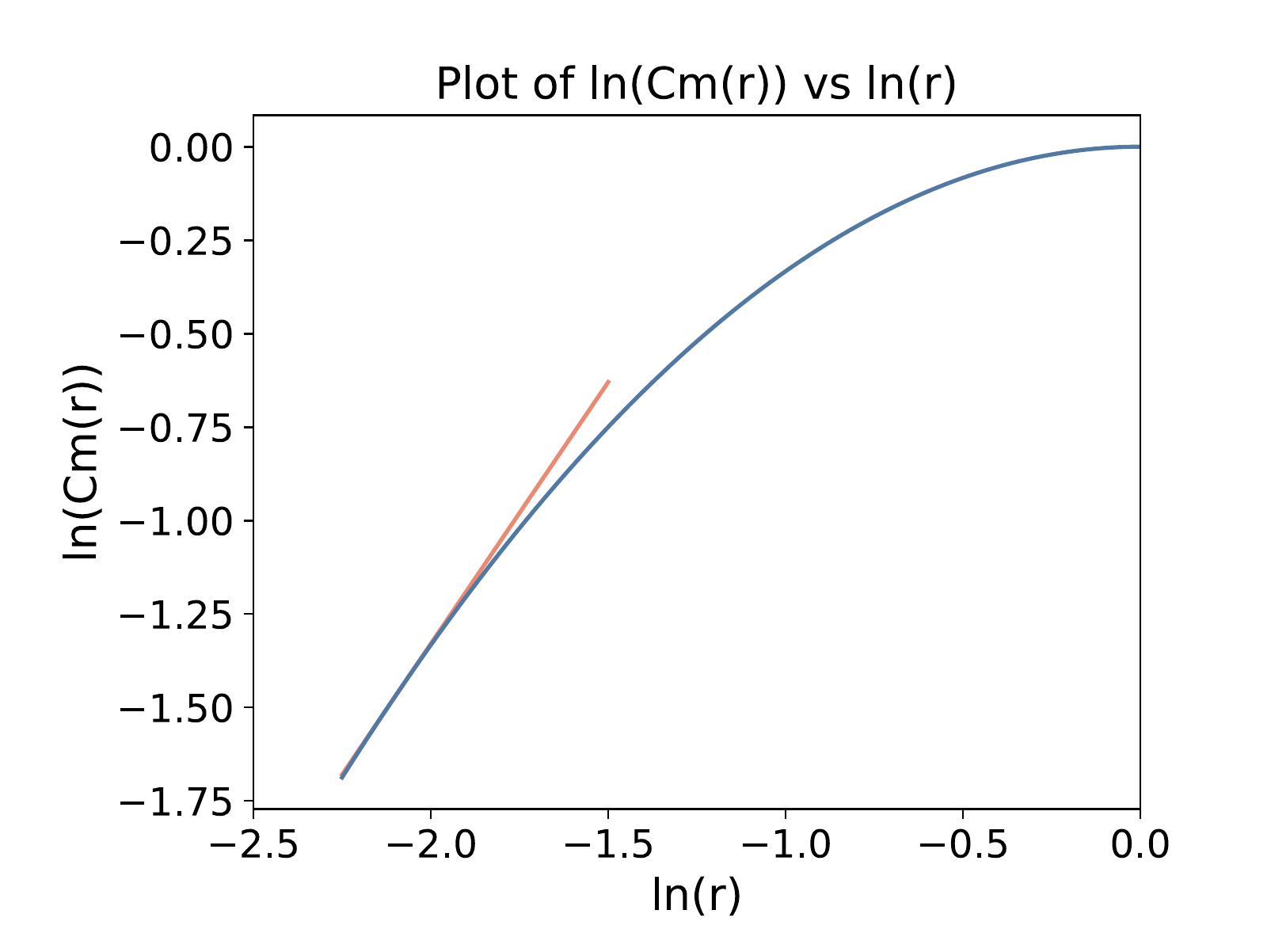}}
\caption{Intrinsic dimensionality
is the maximum slope of the smoothed \textcolor{blue}{blue} curve of 
 $\mathit{ln(r)}$ vs  $\mathit{ln}(C(r))$
 (see the  \textcolor{orange}{orange} line). }\label{fig:dimension}
\end{figure}

For example, in Figure~\ref{fig:dimension}, the intrinsic dimensionality of \textcolor{blue}{blue} curve is 1.6 approximated by the maximum slope which is the \textcolor{orange}{orange} line.

\begin{algorithm} 
\scriptsize
% \scriptsize
\begin{algorithmic}
\SetAlgoLined
\STATE{Import data from \texttt{Testdata.py}}
\STATE{Input: $\mathit{sample}\_\mathit{num}=n, \mathit{sample}\_\mathit{dim}=d$}

    \STATE{$Rs_{\mathit{log}} = \mathit{start:end:step}$}
    \STATE{ $Rs= \mathit{np}.\mathit{exp}(Rs_{\mathit{log}})$}
    \FOR{$R$ in $Rs$}
        % \STATE{\# increasing the Euclidean radius}
        \STATE{\# Calculated by L1 Distance}
        \STATE{$I = 0$}
        \STATE{\# count for pairwise samples within R}
        \FOR{$i,j$ in $\mathit{combinations}(\mathit{data},2)$}
            \STATE{$d=\mathit{distance}(i,j)$}
            \STATE{\# L1 distance}
            \IF{$d<R$}
                \STATE{$I \gets I+1$}
            \ENDIF
        \ENDFOR
        \STATE{$Cr=2*I/n*(n-1)$}
    \ENDFOR
    \STATE{$Crs.\mathit{append}(Cr)$}
    
    \FOR{$i$ in $\mathit{step}$}
    \STATE{$\mathit{gradient}=(Crs[i]-Crs[i-1])/(R[i]-R[i-1])$}
    \STATE{$GR.\mathit{append}(\mathit{gradient})$}
    \ENDFOR
    \STATE{$\mathit{Smooth}(GR)$}
    \STATE{\# smooth the curve}
    \STATE{$\mathit{intrinsicD} \gets \mathit{max}(GR)$}
    \STATE{\# Estimate the intrinsic dimensionality}
\end{algorithmic}
\caption{Intrinsic Dimension by Box-counting Method}
\label{alg:intrinsic}
\end{algorithm}

\begin{figure}
    \centering
    \subfigure[d=5, s=1000]{\includegraphics[width=0.37\textwidth, height=3cm,trim = {0 0 1.6cm 0.8cm}, clip]{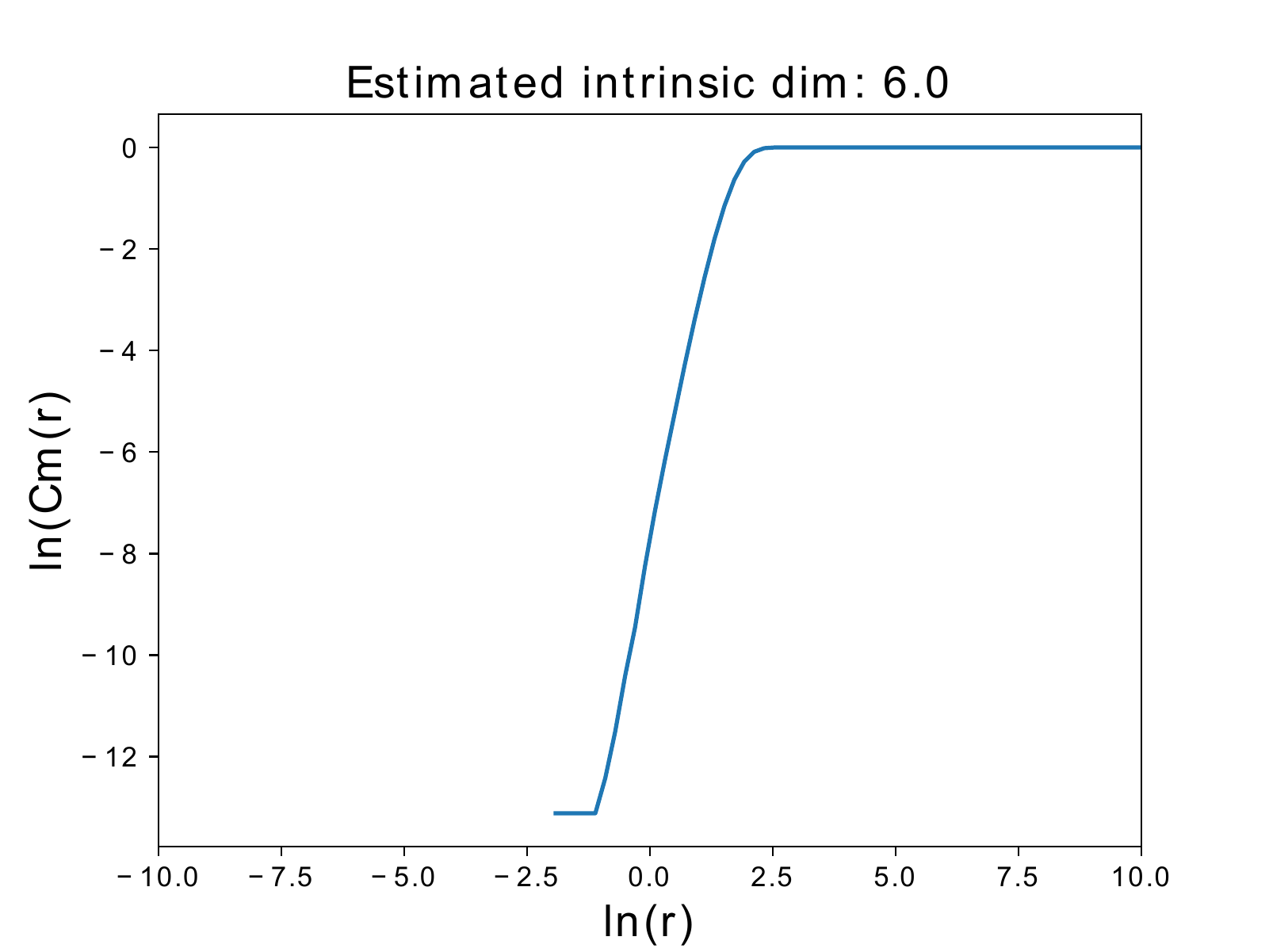}} 
    \subfigure[d=10, s=1000]{\includegraphics[width=0.37\textwidth, height=3cm,trim = {0 0 1.6cm 0.8cm}, clip]{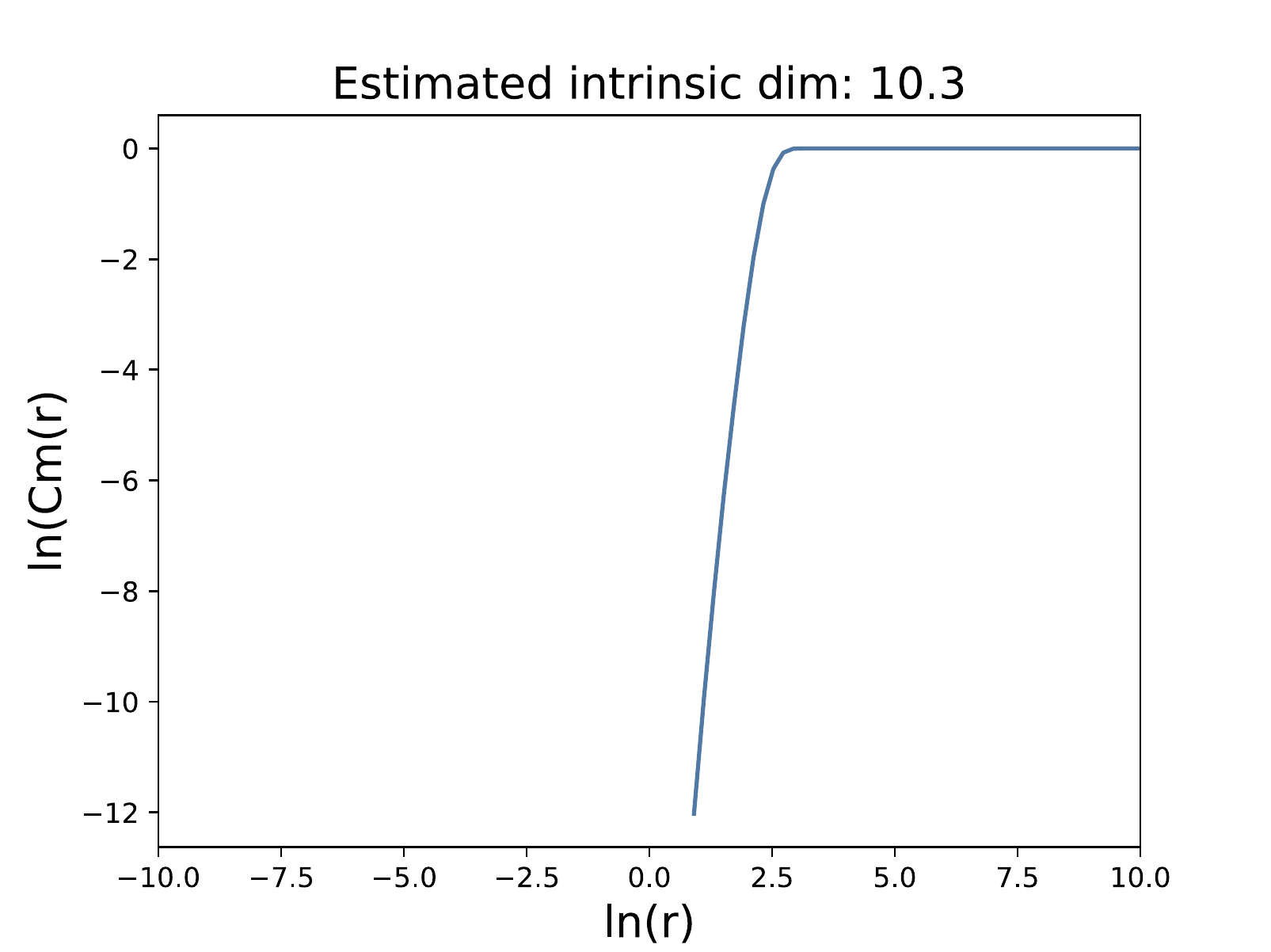}}
    \subfigure[d=20, s=1000]{\includegraphics[width=0.37\textwidth, height=3cm,trim = {0 0 1.6cm 0.8cm}, clip]{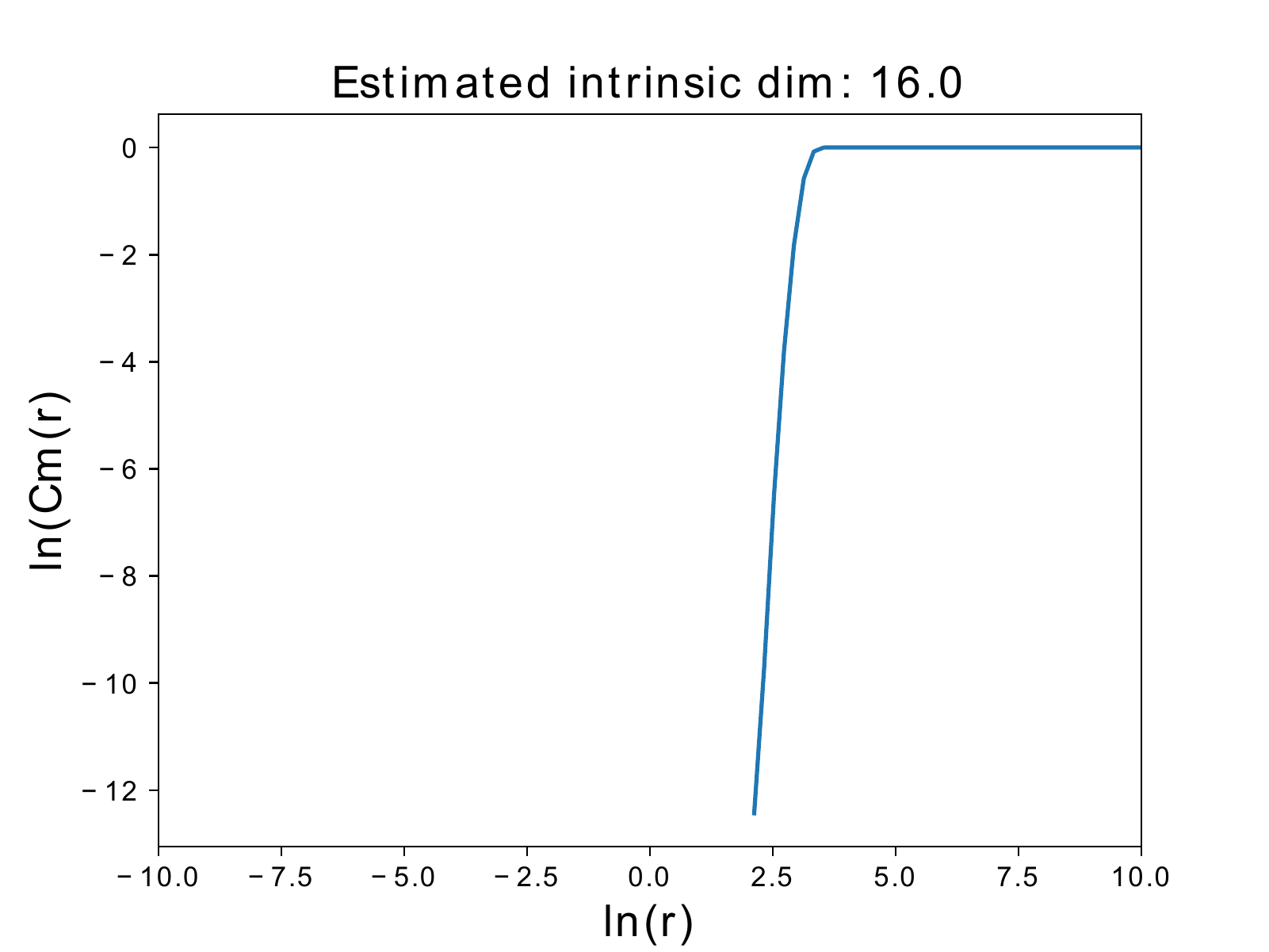}}
    \subfigure[d=40, s=1000]{\includegraphics[width=0.37\textwidth, height=3cm,trim = {0 0 1.6cm 0.8cm}, clip]{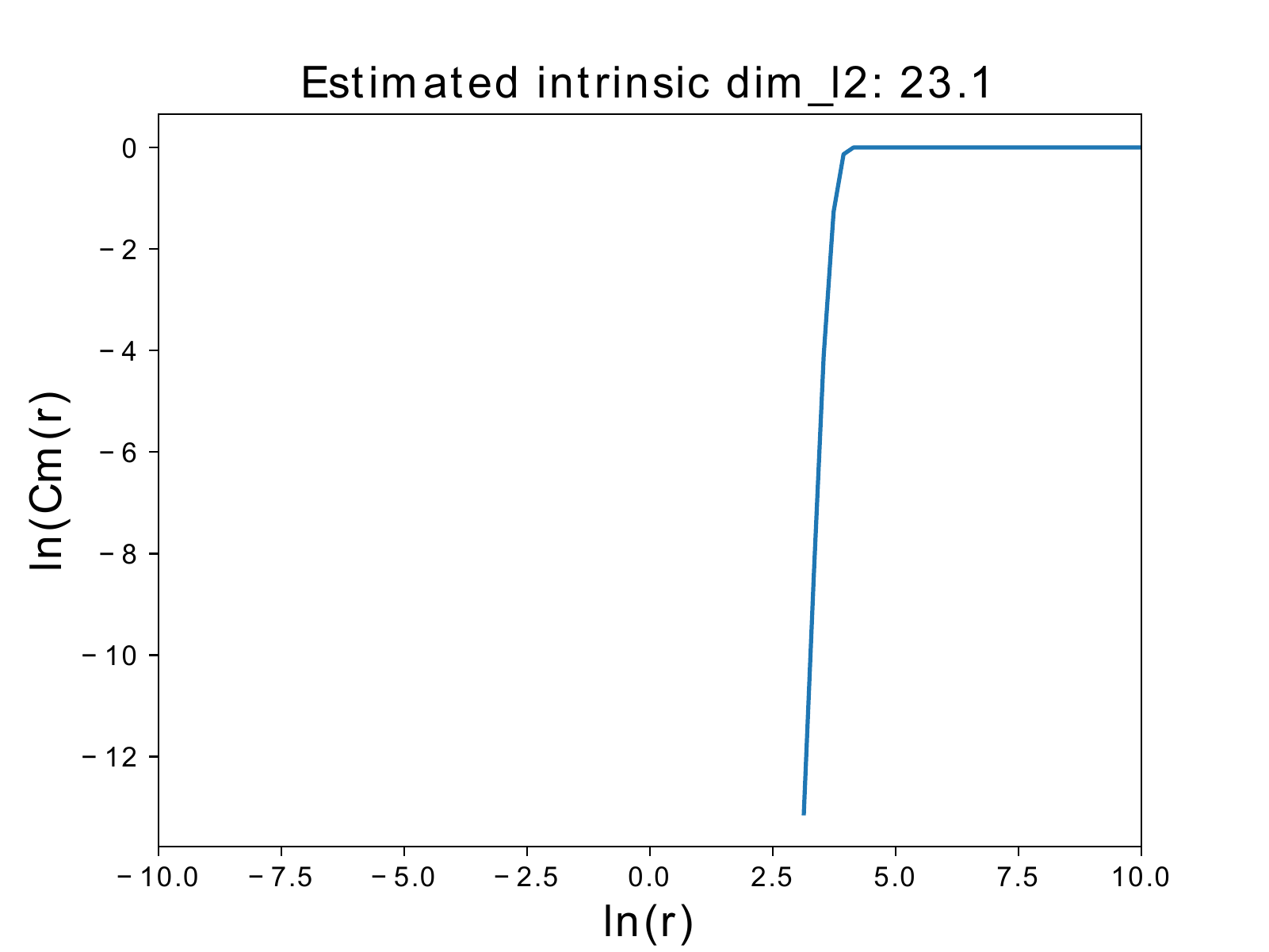}}
    \caption{Algorithm~\ref{alg:intrinsic}  works well for up to 20 intrinsic dimensions. To show that,  
we randomly filled 1000 rows of tables of data with $d$ columns with random variables $0\le X \le 1$.
 Algorithm~\ref{alg:intrinsic} came  close to the actual value of  $d$ for $d<20$. Above that point, the algorithm, seems to underestimate the number of columns-- an effect we attribute to the ``shotgun correlation effect'' reported by Courtney et al.~\cite{courtney93} in 1993.
They reported that, due to randomly generated spurious correlations,  the correlation between $d$ random variables will increase with $d$.
Hence it is not surprising that in the (e.g.) $d=40$ plot of this figure,
we find less than 40 dimensions.
} \label{fig:random}
\end{figure}

% \begin{figure}[!t]
%     \centering
%     \subfigure[d=5, s=1000]{\includegraphics[width=0.48\textwidth]{curve_dim5_l1.pdf}} 
%     \subfigure[d=10, s=1000]{\includegraphics[width=0.48\textwidth]{curve_dim10_l1.pdf}}
%     \subfigure[d=20, s=1000]{\includegraphics[width=0.48\textwidth]{curve_dim20_l1.pdf}}
%     \subfigure[d=40, s=1000]{\includegraphics[width=0.48\textwidth]{curve_dim40_l1.pdf}}
%     \caption{Algorithm~\ref{alg:intrinsic}  works well for up to 20 intrinsic dimensions. To show that,  
% we randomly filled 1000 rows of tables of data with $d$ columns with random variables $0\le X \le 1$.
%  Algorithm~\ref{alg:intrinsic} came  close to the actual value of  $d$ for $d<20$. Above that point, the algorithm, seems to underestimate the number of columns-- an effect we attribute to the ``shotgun correlation effect'' reported by Courtney et al.~\cite{courtney93} in 1993.
% They reported that, due to randomly generated spurious correlations,  the correlation between $d$ random variables will increase with $d$.
% Hence it is not surprising that in the (e.g.) $d=40$ plot of this figure,
% we find less than 40 dimensions.
% } \label{fig:random}
% \end{figure}

Algorithm~\ref{alg:intrinsic}  shows the intrinsic dimensionality calculator used in this paper. Note that this calculator uses  Equation~\ref{eq:cr}
with an L1-norm. 
Figure~\ref{fig:random} displays a verification study which shows that this algorithm works well  for  up to 20 intrinsic dimensions.

\subsection{Intrinsic Dimensionality and Static Code Warnings}
\label{tion:instrinsic_results}
\tbl{dimensionality_L1} shows the results of applying our intrinsic dimensionality calculator   to the static code warning data. In that table, we  observe that:
\bi
\item The size of the data set is not associated with intrinsic dimensionality. Evidence: our largest data set (Lucene) has the lowest intrinsic dimensionality.
\item The intrinsic dimensionality of our data is very low (median value of less than one, never more than two), which is latent dimensionality instead of subsetting from original features.
\ei

% actionable and unactionable samples (denoted as the yellow and purple dots in scatter plots) are mixed up after projecting cass dataset into 2D space with PCA. 
% This indicates ineffectiveness of PCA in this case study. T-SNE~(t-distributed stochastic neighbor embedding )~\cite{maaten2008visualizing} is a non-linear probability-based visualization tool which models each high-dimensional dataset by a two or three-dimensional space in such a way that similar samples are modeled by nearby neighbors and dissimilar samples are modeled by distant neighbors with high probability. This tool is also leveraged in word embedding visualization for the software engineering domain~\cite{efstathiou2018word}. In this work, we utilize T-SNE for visualizing the data with our estimated dimensionality as prior knowledge. After conducting a scatter plot of the mapped dataset, GMM clustering~(denoted as blue ellipses where darker ellipse is area closer to the cluster centroid) is leveraged to depict the distribution pattern of actionable warnings~(denoted as purple dots) and unactionable ones~(denoted as yellow dots). GMM~(Gaussian mixture clustering) is a more robust clustering algorithm than K-means due to its capability to incorporate the uncertainty of the label of incoming samples and group them into cohesive components which are approximate representatives of real patterns within the data set~\cite{rasmussen1999infinite}. Analysis is illustrated in Fig~\ref{fig:visualization}.}

This paper is not the first to suggest that several SE data sets are low dimensional in data.
Menzies et al. also review a range of strange SE results, all of which indicate that the effective number of dimensions of SE data is very low~\cite{menzies07}. 
Also, Agrawal et al.~\cite{agrawal2019dodge} argued that dimensionality of the space of performance  scores generated from some software effectively divides into just a few dozen regions-- which is a claim we could restate as that space is  effectively low dimensional. 
Further, Hindle et al.~\cite{Hindle:2016}  made an analogous argument that:
\begin{quote}
``Programming languages, in theory, are complex, flexible and powerful, but the programs that real people actually write are mostly simple and rather repetitive, and thus they have usefully predictable statistical properties that can be captured in statistical language models and leveraged for software engineering tasks.''
\end{quote}
That said, Hindle, Agrawal, and Menzies et al. only show that there can be a benefit in exploring SE data with tools that exploit low dimensionality. None of that work makes the point made in this paper, that for SE data it can be harmful to explore low dimensional SE data with tools designed for synthesizing models from high dimensional spaces (such a deep learners).

 \begin{table}
\caption{Summary of dimensionality of nine datasets.
Calculated using Equation~\ref{eq:cr}.}
% \small
\footnotesize
% \begin{adjustbox}{max width=0.48\textwidth}
\begin{tabular}{@{}cccc@{}}
\toprule
\multicolumn{1}{l}{}  Dataset & \begin{tabular}[c]{@{}c@{}}original\\ dimensionality\end{tabular} & \begin{tabular}[c]{@{}c@{}}intrinsic \\ dimensionality\end{tabular} & \begin{tabular}[c]{@{}c@{}}instance \\ counts\end{tabular} \\ \midrule

  lucence & 57 & \colorbox[HTML]{F79B00}{0.15} & 3259 \\
  phoenix & 44 & \colorbox[HTML]{F79B00}{0.62} & 2235 \\
  tomcat & 60 & \colorbox[HTML]{F79B00}{0.73} & 1435 \\
  derby & 58 & \colorbox[HTML]{F79B00}{0.78} & 2479 \\
  Ant & 56 & \colorbox[HTML]{F79B00}{0.82} & 1229 \\
 commons & 39 & \colorbox[HTML]{F79B00}{1.04} & 725 \\ 
  mvn & 47 & \colorbox[HTML]{F79B00}{1.10} & 813 \\
  jmeter & 49 & \colorbox[HTML]{F79B00}{1.54} & 604 \\
  cass & 55 & \colorbox[HTML]{F79B00}{1.94} & 2584 \\\bottomrule
\end{tabular}
% \end{adjustbox}
\label{tbl:dimensionality_L1}
\end{table}

\begin{figure}[!t]
    \centering
    \subfigure[Cass: Estimated Dim = 1.94, visualized in 2-D space via PCA]{\includegraphics[width=0.45\textwidth]{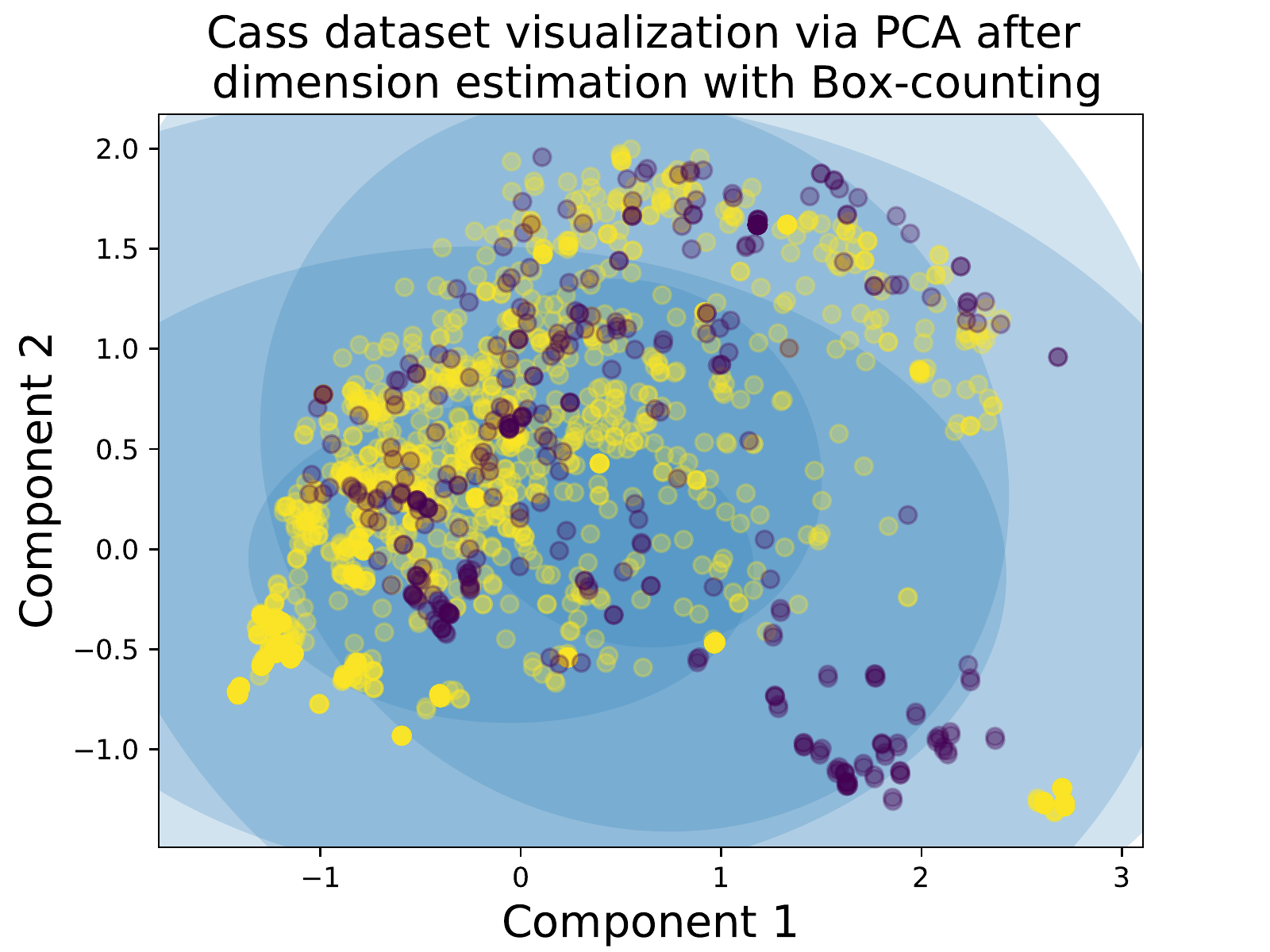}} 
    \subfigure[Cass: Estimated Dim = 1.94, visualized in 2-D space via T-SNE]{\includegraphics[width=0.45\textwidth]{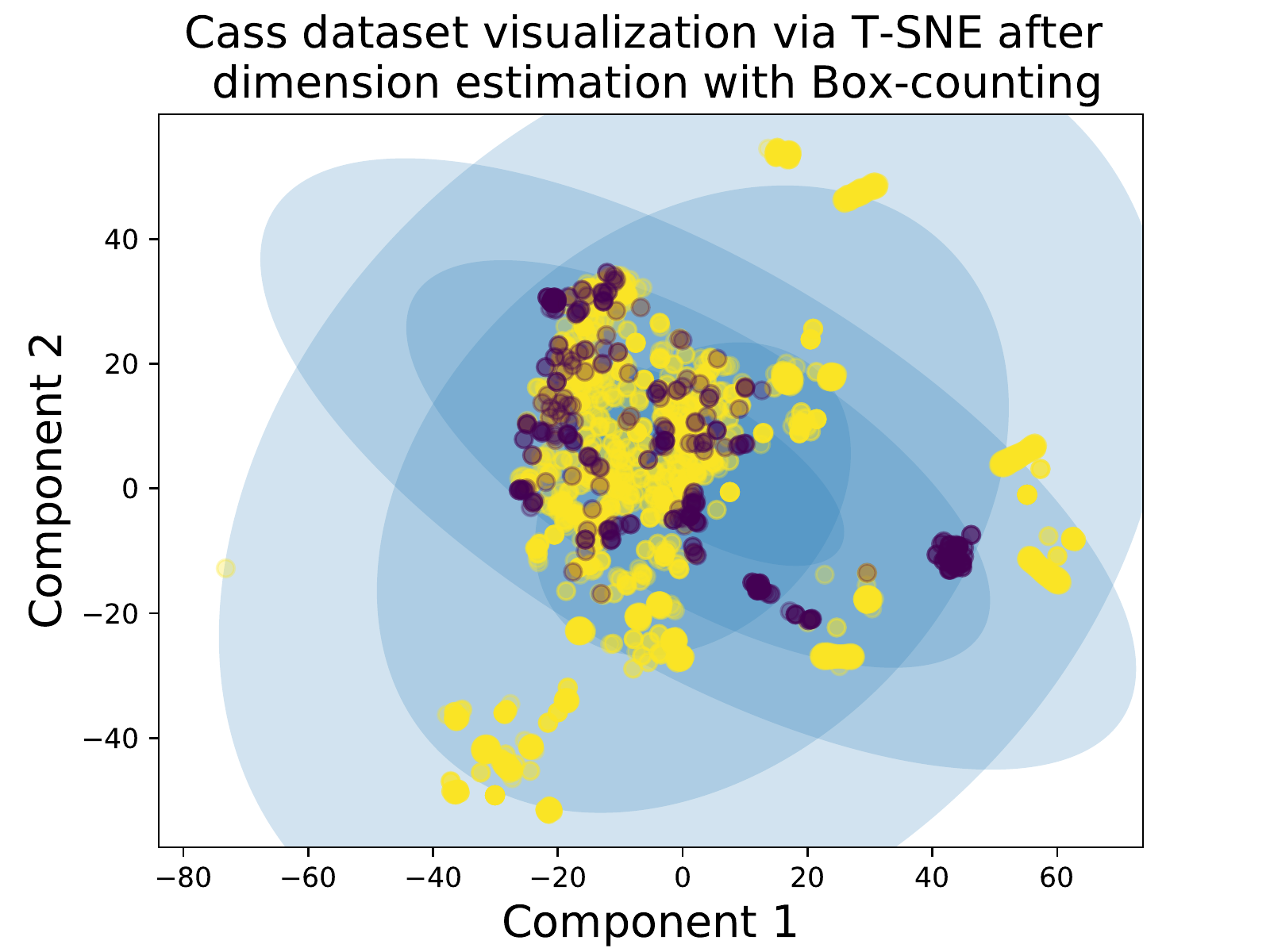}}
    \subfigure[Commons: Estimated Dim = 1.04, visualized in 2-D space via PCA]{\includegraphics[width=0.45\textwidth]{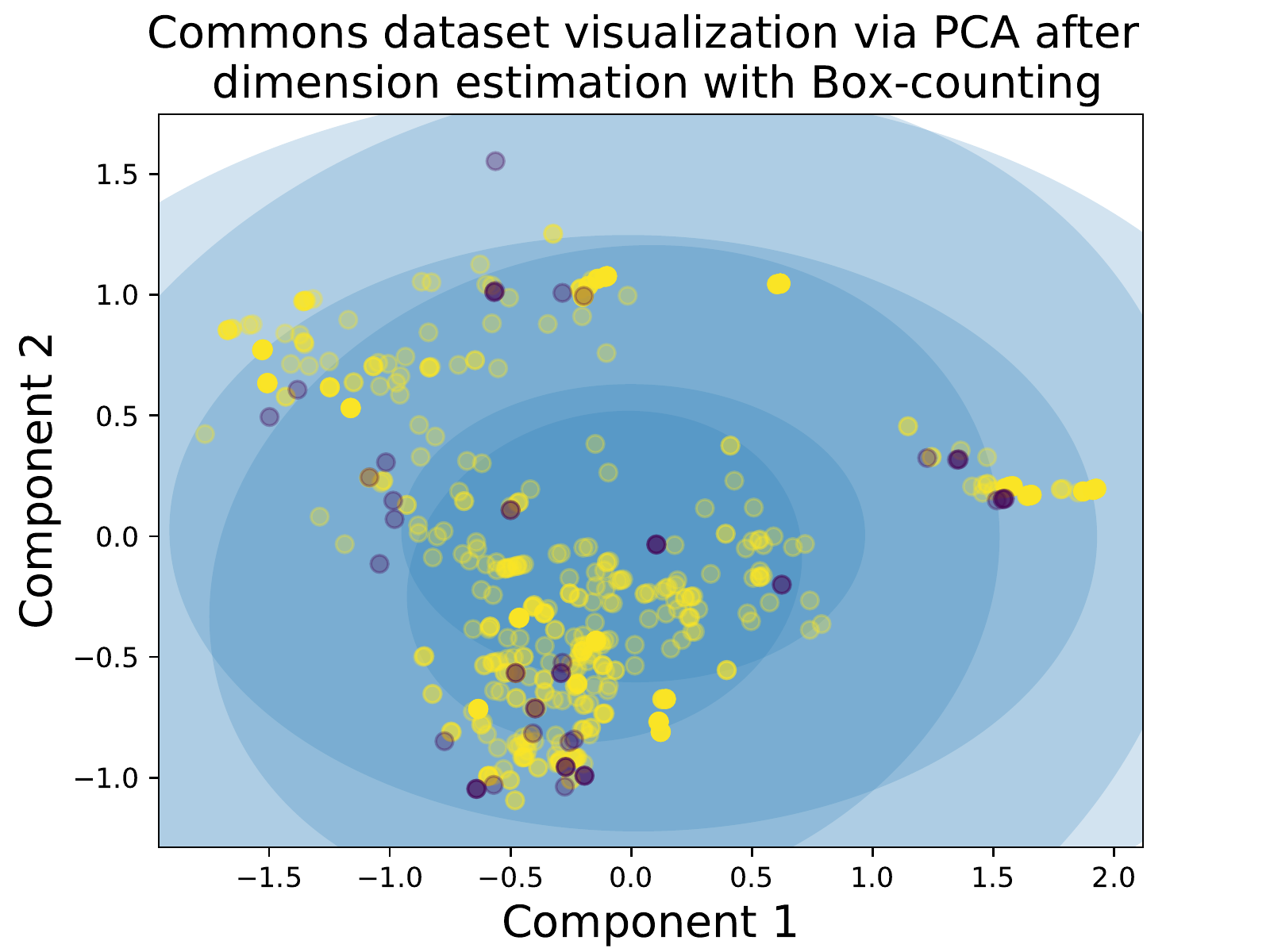}} 
    \subfigure[Commons: Estimated Dim = 1.04, visualized in 2-D space via T-SNE]{\includegraphics[width=0.45\textwidth]{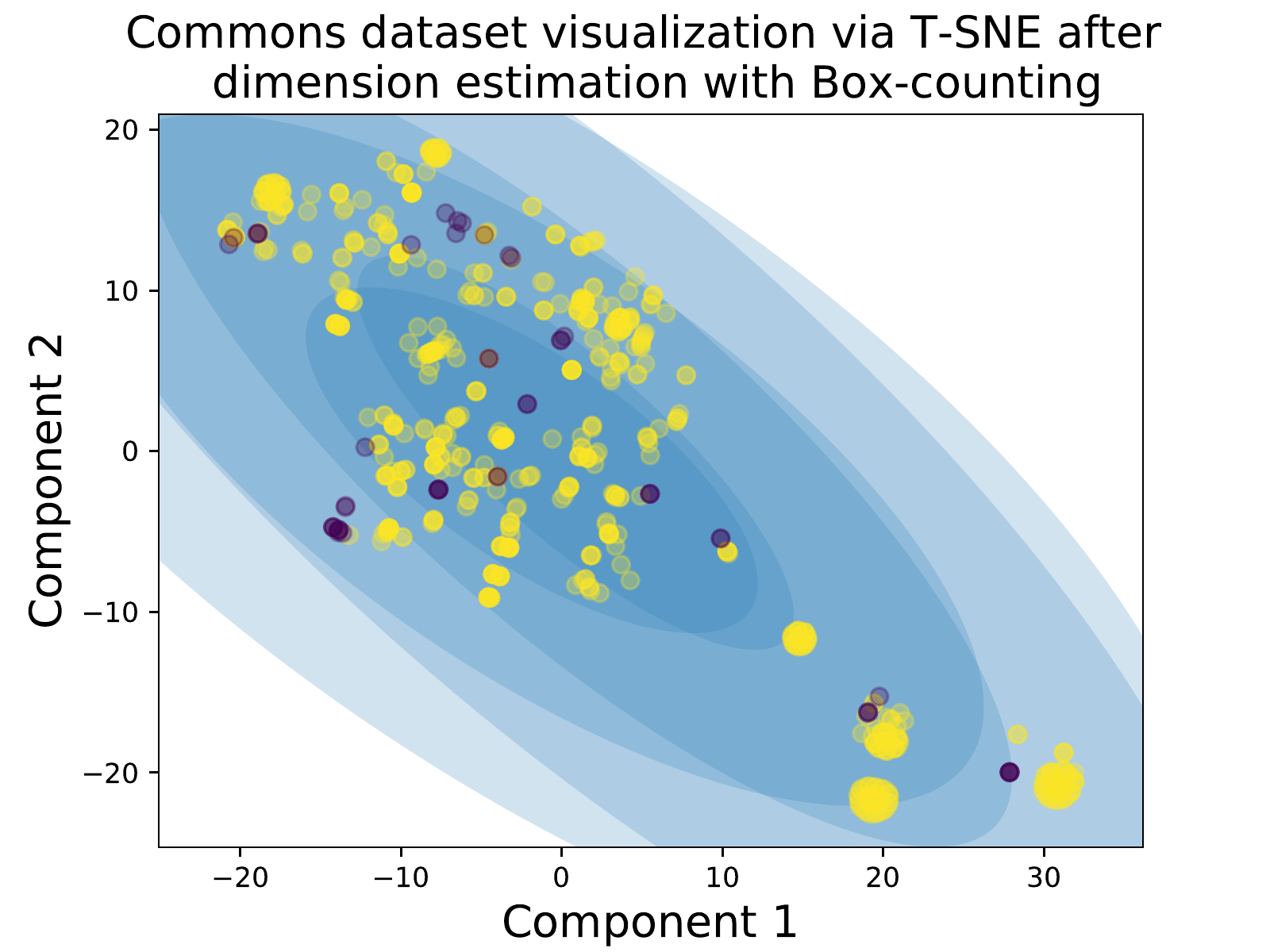}}
    \caption{In this figure, we tried visualizing the underlying intrinsic dimensions of our data using PCA and T-SNE. In summary, those tools were not useful for finding those dimensions. We say that since  the clusters shown here do not separate the actionable and unactionable warnings. 
    % sumamry Data visualization. Cass and Commons datasets are explored in visualization because their intrinsic dimensionality is estimated as 1.94 and 1.04 which are close to integers 2 and 1 so that the visualization can be well depicted in a visual 2-D space.  
    Results here come from GMM~(Gaussian mixture clustering). This algorithm is known to be more robust than K-means due to its capability to incorporate the uncertainty of the sample labels~\cite{rasmussen1999infinite}. In these figures:
    (a)~the x-y plots  from the first two components found by PCA and T-SNE;
    (b)~darker ellipses are regions closer to the cluster centroids;
    (c)~actionable warnings are shown in purple dots;
    (d)~unactionable warnings are shown in yellow. Data from these figures come from cass and commons. Results from other data sets are just as uninformative. In future work, it could be useful to return to this goal of trying to visualize the underlying intrinsic dimensions, perhaps with other clustering tools.
} \label{fig:visualization}
\end{figure}

Lastly, we report on a failed attempt
to visualize our data.  
Recall from the above that Levina et al. caution against the
use  PCA.  Consistent with
that warning, we found PCA plots to be uninformative. We further incorporate a nonlinear visualization tool (T-SNE~\cite{maaten2008visualizing}), a widely used probability-based model to depict our data set. As shown in  Figure~\ref{fig:visualization},
those tools were not useful for separating out our classes. In future work, it could be useful to return to this goal (of trying to visualize the underlying intrinsic dimensions), perhaps with other clustering tools.

 \subsection{Summary}
After applying Algorithm~\ref{alg:intrinsic} to our data, we can assert that static code warning is inherently a low dimensional problem.
Specifically, our data sets can be characterized with less than two dimensions as reported in \tbl{dimensionality_L1}.
% satisfies the third condition to demonstrate Camastra's Law. 
Hence, we believe that the reason deep learning performs so similarly or even worst than conventional learners for static code warnings is that it is a very big hammer being applied to a very small nail.

\section{Discussion}
\label{sec:discussion}

\subsection{Threats to Validity}
 
 Before discussing threats to validity, we make the meta-comment that induction is {\em not} a certain inference since (e.g.) just because the sun has risen every day since the birth of this planet, we cannot say with 100\% certainty that it will rise tomorrow. Project data differs from project to project and just because we found that past data generated  a clear signal on what was an actionable code warning (see \tbl{summary}), there is no guarantee that such a signal will be found in future data. Hence, any data mining paper will raise issues of (e.g.) sampling bias and measurement bias.~\\ 
\hspace{5mm}To address this concern,
we assert that data mining researchers need to express their conclusions in a {\em refutable} manner. To that end, 
we take care to draw our conclusions from data and tools that are free to download\footnote{\url{https://github.com/XueqiYang/intrinsic_dimension}} and which are distributed under open source licenses that enable their widespread use.

\textbf{Sampling bias.}
In terms of sampling bias, our first comment is that all our conclusions are based on the data explored in the above experiments.
For future work, we need to repeat this analysis using different data sets. Our second comment is while we depreciate deep learning, that warning only applies to low dimensional data. Deep learning is very useful for very high dimensional problems; e.g. vision systems in autonomous cars.   

For this specific research task, we explore the static warning data set collected by a prior EMSE paper. Wang's study reduces our sampling bias issue (due to the thoroughness of that analysis). That said, sample bias threatens any paper on analytics (not just this one) since conclusions that hold for one project may not hold for any other. No paper can explore all data sets -- the best we can do (and we have done) is carefully documenting our methods and placing our tools on a repository  that others can access (so the community can easily apply our methods to their data).

% \textbf{Optimizer bias.}
% This work applies Adam optimizer in the training process of deep neural network~\cite{kingma2014adam}. Adam is an efficient accelerator to facilitate DNN models. There are much more optimizer, like Stochastic Gradient Descent (SGD) which has been extensively utilized to search optimal solutions of deep learning models in high-dimensional parameter space. However, SGD is also denounced due to its weakness, where slow training and convergence makes it computationally expensive and the inherently sequential feature makes it difficult to parallelize the computation using GPUs or distributed computer clusters. We use Adam as solver to optimize the training process which successes the merits of two currently popular methods: AdaGrad and RMSProp, which works well for sparse gradients with a fast convergence speed.

% However, applying our model to other SE tasks doesn't necessarily assert the best performance before considering other deep learning optimizer.
% However, this doesn't necessarily guarantee a best performance in other domains or other static warning datasets. According to 
% the No Free Lunch Theorems~\cite{wolpert1997no}, applying our method framework to other areas would be needed
% before we can assert that our methods are also better in those domains.

% And reckless conclusions should not be drawn if few of these models obtain bad performance.

\textbf{Measurement bias:}
To evaluate the efficiency
% validity
of our learners, we employ three commonly used measurement metrics in SE area: recall, false alarm, and AUC. Several prior research works have demonstrated the necessity and effectiveness of these measurements~\cite{yu2018finding,wang2018there}. For the static warning analysis task, the major goal is to identify the true positives, so recall and false-positive rate are adequate measurements. Also, AUC indicates the overall performance of a classifier. 
% Several prior research work has demonstrated the necessity and effectiveness of these measurements.
There exist many other metrics widely adopted by SE community, like F1 score, G measure and so forth. For the same research question, different conclusions may be drawn by using various evaluation metrics. In future work, we would use other evaluation metrics to have a more comprehensive analysis.

% many studies are still based on some classic and traditional metrics, eg. confusion matrix or also known as error matrix. There exist many popular terminology and derivations from confusion matrix, false positive, F1 score, G measure and so on. We cannot explore and include all the options in one article. Also, even for this same research methodology, conclusions drawn from different evaluation matrix may differ. However, in this research scenario, this more efficient to report recall and cost for effort-aware model.

\textbf{Parameter bias:} This paper  used 
the default settings for our learners (exception: we adjusted the number of epochs used in our deep learners).
Recent work~\cite{Tantithamthavorn16,agrawal2018better,agrawal2019dodge} has shown that these defaults can be improved via hyperparameter optimization (i.e., learners applied to learners to learn better settings for the control parameters).
In this study, we found that even with the default parameters we could outperform deep learning and prior state-of-the-art results~\cite{wang2018there}. Hence, we leave hyperparameter optimization for future work.

\textbf{Learner bias.}
One of the most important threats to validity is learner bias, since there is no theoretical reason that any learner outperforms others in all test cases.
Wolpert et al.~\cite{wolpert1997no} and Tu et al. \cite{Tu18Tuning} proposed that no learner necessarily works better than others for all possible optimization problems. Moreover, there also exist many other deep models developed in deep learning revolution. Different models show significant advantages in different tasks. For instance, LSTM is utilized in Google Translate to translate between more than 100 languages efficiently, while CNN is widely used in tasks of analyzing visual imagery. In this case, researchers may find other deep neural networks work better on SE tasks. For future work, we need to repeat this analysis using different learners.

\subsection{Future Work}

In future work, it would be interesting to do more comparative studies of SE data using  deep learning versus other kinds of learners. Those studies should pay particular attention to the issue raised here; i.e. does DL match the complexity of datasets in other SE areas? Also, we can exploit the deep learning effect
described above to generate a new generation of better learners.
In the literature, non-linear mapping methods that can project complex features into lower dimension space are widely explored in the areas of statistics and computer vision~\cite{krizhevsky2012imagenet}. Such feature reduction can significantly save computational overhead brought by complex algorithms such as DNN models. Therefore, the implementation of non-linear feature mapping might dispel the concern of SE researchers
% \revised{and practitioners }
caused by the overwhelming running cost of deep learning models on big datasets 
(as well as contribute to the promotion of deep learning in SE area). A comprehensive implementation of non-linear feature mapping is left to future work.

Also, another avenue for future work would be to drill down into specific warnings types for static warnings generated by Findbugs. As illustrated in prior literature~\cite{khalid2015examining}, more than 400 possible types of warnings identified by Findbugs can be categorized into eight groups  (bad practice, correctness, internationalization, malicious code vulnerability, multi-threaded correctness, performance, security, and dodgy code). These warnings can be assigned a priority (e.g., low, medium, and high) which indicates how confident this warning is an actionable one or true positive. We think it would be insightful to check how specific types of warnings are essential for the software community
% \revised{and practitioners }
because prior literature indicated specific types of  warnings (e.g., bad-practice, internationalization and, performance) had significantly higher densities in low-rated apps than high-rated ones. Such a more ``fine-grained'' investigation could provide a very meaningful guideline for SE researchers.
% software practitioners.

In addition, we plan to apply these static analysis methods to other kinds of data. Currently, we are in open discussions with the Linux developers about applying these methods to their domain for the particular task of reducing the number of false alarms generated by security static code analyzers. This work is in progress and, to date, we have nothing definitive to report.

Finally, in Figure~\ref{fig:visualization} we show failed experiments in trying to visualize the underlying intrinsic dimensions. In future work, it would be useful to return to this goal, perhaps with other clustering tools.

\section{Conclusion}
\label{sec:conclusion}

% Some researchers are worried that DL is being applied with insufficient critical assessment. 
% For example, Gary Marcus~\cite{marcus2018deep} 
% warns than DL may be so over-hyped that it runs  ``fresh risk for seriously dashed expectations'' that could blinker AI researchers from trying new ideas as well as
% bringing  another AI winter (the period in the mid to late 1980s when most AI companies went bankrupt due to poor generalization
% of their methods). 

Static code analysis tools produce many false positives which many programmers ignore.
Such tools can be augmented with data mining algorithms to prune away the spurious reports, leaving behind just the warnings that cause programmers to take action to change their code. As seen by the above results, such data miners can be remarkably effective (and exhibit very low false alarm rates, very high AUC results, and respectably high recall results).

In this paper, we perform an empirical experiment to apply tree learners, linear SVM, and deep learning (with early stopping) to predicting actionable static warning analysis tasks on nine software projects. 
% We find that deep learning appeared to work best-- until a seemingly small change in the test rig revealed that that learner was overfitting to the data. 
We find deep learners mismatch the complexity of our static warning datasets with high running cost. Using a dimension reduction algorithm, our static warning datasets are reported as inherently low dimensional. As suggested by Principle of Parsimony, it is detrimental to employ sophisticated models (like deep learning) on data that is inherently low dimensional (like the data explored here).
Hence, we endorse the use of linear SVM for predicting which static code warnings are actionable.

To end, we note the irony of this paper. In this paper,   we turned to intrinsic dimensionality {\em after} exploring  particular data set-- which is the {\em opposite} for what we  are recommending for future practice. 
For future work in software analytics, we suggest that analysts match the complexity of their analysis tools to the underlying complexity of their research problem. Specifically, we strongly urge the SE community to compute the dimensionality of their data, then use those results to select an appropriate analysis algorithm.

\section{Acknowledgment}
This work was partially funded by an NSF award
\#1703487.

\balance

% BibTeX users please use one of
% \bibliographystyle{spbasic}      % basic style, author-year citations
\bibliographystyle{spmpsci}      % mathematics and physical sciences
%\bibliographystyle{spphys}       % APS-like style for physics
%\bibliography{}   % name your BibTeX data base

% \bibliographystyle{ACM-Reference-Format}
\bibliography{references}
% \bibliography{main}

\clearpage
\setcounter{page}{1}
\pagenumbering{roman}
\normalsize
\end{document}